\documentclass[aps, prx,english,a4paper,floatfix,twocolumn,longbibliography,groupedaddress, superscriptaddress]{revtex4-2}

\usepackage{amsmath}
\usepackage{color}
\usepackage{tikz}
\usetikzlibrary{quantikz2}
\usepackage{graphicx, import}
\usepackage{hyperref}
\usepackage{parskip} 
\usepackage{glossaries}
\usepackage{booktabs}
\usepackage{multirow}
\usepackage{soul}
\usepackage[normalem]{ulem}
\usepackage{algorithm}
\usepackage{algpseudocode}
\usepackage{placeins}
\usepackage{lipsum}
\usepackage[dvipsnames]{xcolor}
\usepackage{braket}
\usepackage[final]{pdfpages}

\makeatletter
\AtBeginDocument{\let\LS@rot\@undefined}
\makeatother

\usepackage{xr}
\makeatletter

\newcommand*{\addFileDependency}[1]{
    \typeout{(#1)}
    \IfFileExists{#1}{}{\typeout{No file #1.}}
}\makeatother

\newcommand*{\myexternaldocument}[1]{%
    \externaldocument{#1}%
    \addFileDependency{#1.tex}%
    \addFileDependency{#1.aux}%
}

\myexternaldocument{supp}

\begin{document}
\newacronym{aqc}{AQC}{approximate quantum compiling}
\newacronym{mps}{MPS}{matrix product state}
\newacronym{dmrg}{DMRG}{density matrix renormalisation group}
\newacronym{bahc}{BAHC}{bond-alternating Heisenberg chain}
\newacronym{spt}{SPT}{symmetry-protected topological}
\newacronym{zne}{ZNE}{zero-noise extrapolation}
\newacronym{trex}{TREX}{twirled readout error extinction}
\newacronym{aklt}{AKLT}{Affleck-Kennedy-Lieb-Tasaki}
\newacronym{el}{EL}{exact ladder}
\newacronym{al}{AL}{approximate ladder}
\glsdisablehyper

\title{Preparing 100-qubit symmetry-protected topological order on a digital quantum computer}

\author{George Pennington}
\affiliation{The Hartree Centre, STFC, Sci-Tech Daresbury, Warrington WA4 4AD, UK}
\author{Kevin C. Smith}
\affiliation{IBM Quantum, IBM Research Cambridge, Cambridge, MA 02142, USA}
\author{James R. Garrison}
\affiliation{IBM Quantum, Thomas J. Watson Research Centre, Yorktown Heights, NY, USA}
\author{Lachlan P. Lindoy}
\affiliation{National Physical Laboratory, Hampton Road, Teddington TW11 0LW, United Kingdom}
\author{Jason Crain}
\affiliation{IBM Research Europe, The Hartree Centre, Sci-Tech Daresbury, Warrington WA4 4AD, UK}
\affiliation{Department of Physics, Clarendon Laboratory, University of Oxford, Oxford OX1 3QU, UK}
\author{Ben Jaderberg}
\affiliation{IBM Quantum, IBM Research Europe, Hursley, Winchester, SO21 2JN, United Kingdom}

\date{\today}

\begin{abstract}
%TC:ignore
Symmetry-protected topological (SPT) phases extend the Landau paradigm of quantum matter by admitting distinct symmetry-preserving phases that lack any local order parameter. Demonstrating these phases at scale on programmable quantum processors is a key milestone in using quantum hardware to probe emergent many-body phenomena, yet it is impeded by the circuit depth normally required to capture non-trivial entanglement. Here we use a tensor network based approximate quantum compiling (AQC) protocol to construct shallow quantum circuits (18--39 CNOT depth), which prepare 100-site ground states of the spin-$1/2$ bond-alternating Heisenberg chain in both SPT phases, to $97.9$--$99.0\%$ fidelity. Upon executing the circuits on IBM quantum hardware, the resulting states exhibit all defining signatures of SPT order including non-local string order for strings of up to length 20, characteristic degeneracies in the entanglement spectrum and clear evidence of symmetry-protected edge modes. The simultaneous observation of these independent diagnostics establishes current quantum computers as versatile platforms for large-scale studies of symmetry-protected quantum matter. More broadly, our results establish a practical foundation for probing non-equilibrium quench dynamics of such systems in regimes that challenge classical computational methods.
%TC:endignore
\end{abstract}

\glsresetall

\maketitle

\begin{figure*}[ht]
    \centering
    \includegraphics[width=0.8\linewidth]{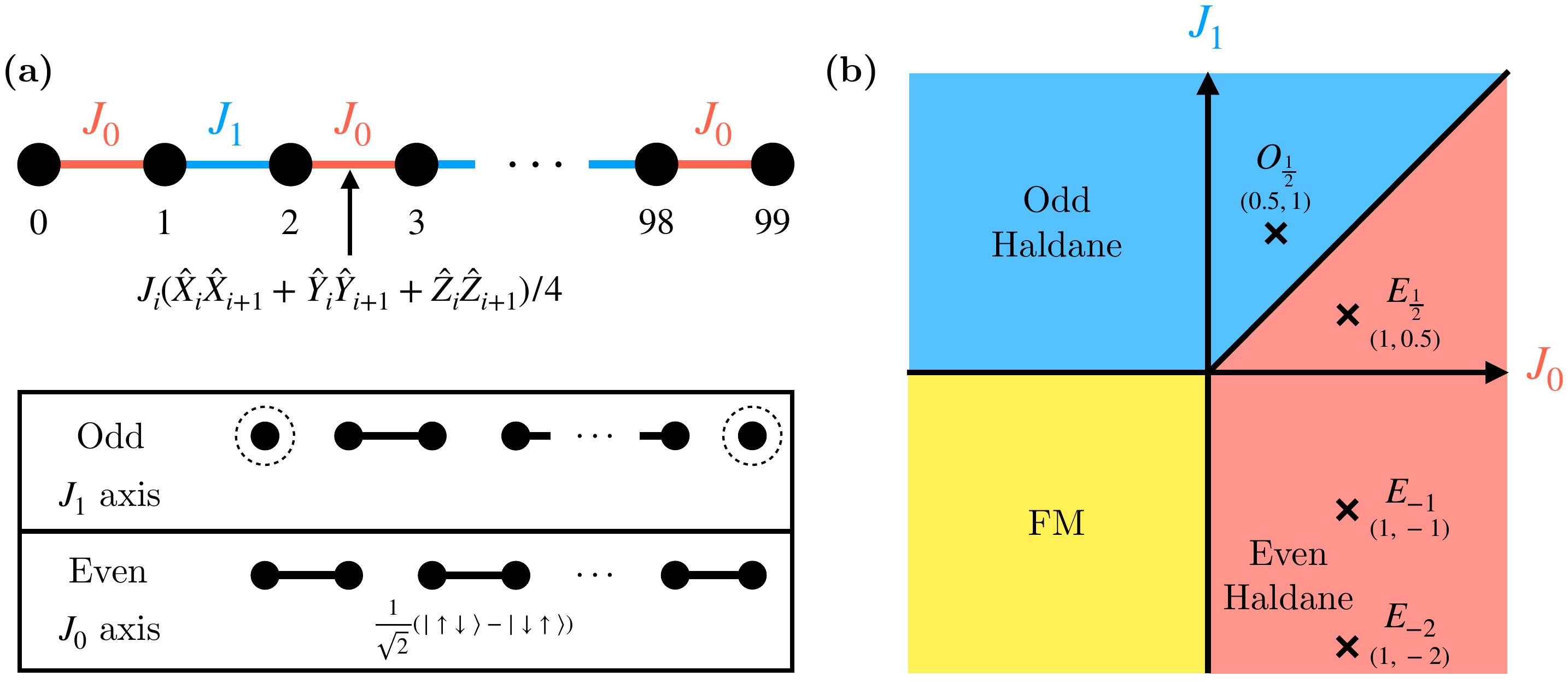}
    \caption{a) (Top) A diagram representing the 100-site spin-1/2 bond-alternating Heisenberg chain, with alternating $J_0$, $J_1$ couplings. (Bottom) a schematic representation of the ground states for the positive $J_0$ and $J_1$ axes, consisting of simple products of singlet pairs: $1/\sqrt{2}(\ket{\uparrow\downarrow}-\ket{\downarrow\uparrow})$, plus decoupled edge spins in the odd-Haldane phase. b) The phase diagram of the model~\cite{haghshenas2014symmetry, wang2013topological}. The model exhibits two \gls{spt} phases: the odd-Haldane phase for  $J_1>0$ and $J_1>J_0$ and the even-Haldane phase for $J_0>0$ and $J_0>J_1$, along with a trivial ferromagnetic phase for $J_0<0$ and $J_1<0$. The black crosses denote the four points in the phase diagram considered in this work, which we denote as $O_\frac{1}{2}$, $E_\frac{1}{2}$, $E_{-1}$, and $E_{-2}$.}
    \label{fig:1_model}
\end{figure*}

The discovery of \gls{spt} phases has transformed our understanding of quantum matter by revealing classifications of order that transcend the Landau paradigm~\cite{senthil2015symmetry}. Unlike conventional phases characterised by local order parameters, \gls{spt} phases preserve the protecting symmetries and are diagnosed by non-local or entanglement-based invariants together with symmetry-protected edge states that are robust to symmetry-preserving perturbations~\cite{pollmann2012string, chen2011classification}. As a result, \gls{spt} phases have become a central topic in modern condensed matter physics~\cite{senthil2015symmetry}. These features not only enrich the taxonomy of quantum phases, but also underpin emerging applications in quantum information science~\cite{miyake2010quantum}, including topologically protected spin transport~\cite{kane2005quantum}, low-dissipation interconnects~\cite{caceres2023edge} and robust memory elements~\cite{roberts2020symmetry}.

\gls{spt} physics has been explored experimentally in a number of platforms including cold atoms~\cite{song2018observation, pimonpan2022realizing}, Rydberg analogue simulators~\cite{de2019observation, mogerle2025spin} and trapped-ion systems~\cite{cohen2015simulating}. In particular, two realisations of the Haldane phase: the spin-$1/2$ \gls{bahc}~\cite{zhao2024tunable} and the spin-1 bilinear-biquadratic (BLQP) model~\cite{mishra2021observation}, have recently been realised in nanographene structures. Nevertheless, such approaches are naturally constrained by the physical couplings and measurements that are accessible in material systems. This motivates the use of digital quantum computers, where \gls{spt} phases of arbitrary Hamiltonians can be prepared, programmable perturbations can probe static and dynamic properties, and arbitrary observables can be measured including non-local string order.

Recent work has begun to demonstrate the promise of digital quantum platforms toward probing nontrivial \gls{spt} physics. For the canonical example of the spin-1 \gls{aklt} state~\cite{affleck1987rigorous}, methodological innovations have enabled efficient preparation of small systems using digital quantum computers~\cite{smith2023deterministic, edmunds2025symmetry}. Ground state preparation in the \gls{spt} phase of the cluster-Ising model has also seen steady progress~\cite{smith2022crossing, herrmann2022realizing, shen2025robust} including for systems as large as 80 sites~\cite{scheer2025renormalization}. Yet preparing ground states beyond minimally-entangled examples, whilst maintaining large system sizes and verifying their preparation to high fidelity, remains an open problem. In particular, the ability to study spin-1/2 models with only two-body exchange interactions would bring digital quantum simulation closer to the microscopic Hamiltonians recently realised in engineered spin materials~\cite{zhao2024tunable}.

In this work, we make major strides toward addressing these challenges by preparing 100-site ground states of both \gls{spt} phases of the spin-1/2 \gls{bahc} on IBM superconducting quantum computers. We verify our preparation through observation of non-local string order for strings of up to length 20, detection of edge modes, and qualitative observation of signature entanglement spectrum degeneracies. We achieve this by applying an \gls{aqc} protocol to obtain shallow quantum circuit representations of \glspl{mps} produced by the \gls{dmrg} algorithm. For four points in the phase diagram, we obtain ground states with up to bond dimension $\chi=40$, and construct circuits of up to 39 CNOT depth whilst maintaining circuit fidelities of at least 97.9\% with the \gls{dmrg} ground states. This enables a major step forward in the preparation of \gls{spt} states on quantum hardware. For example, prior work demonstrated the preparation of 80-site \glspl{mps} \cite{scheer2025renormalization}, yet was limited to just bond dimension 2. Using our compiled circuits, we observe non-local string order across up to 20 sites, evidence of characteristic degeneracies in the entanglement spectrum, and experimental observation of the symmetry-protected edge states, thereby verifying preparation through signature properties of the \gls{spt} order on quantum hardware. Our method establishes a practical foundation for further research requiring high-fidelity ground state preparation, such as the simulation of non-equilibrium \gls{spt} dynamics~\cite{mazza2014out, mcginley2018topology, hagymasi2019dynamical}, or the use of \gls{spt} states as a resource for measurement-based quantum computation~\cite{else2012symmetry}.

\section*{The bond-alternating Heisenberg chain}\label{sec:bahc}

We consider the 1D spin-1/2 \gls{bahc}, defined by the Hamiltonian:

\begin{equation}
    \hat{H} = \frac{1}{4} \sum_{i=0}^{N-2}J_{i}\left(\hat{X}_i\hat{X}_{i+1} + \hat{Y}_i\hat{Y}_{i+1} + \hat{Z}_i\hat{Z}_{i+1}\right),
    \label{eq:bahc_hamiltonian}
\end{equation}

\begin{figure*}[ht]
    \centering
    \includegraphics[width=0.8\linewidth]{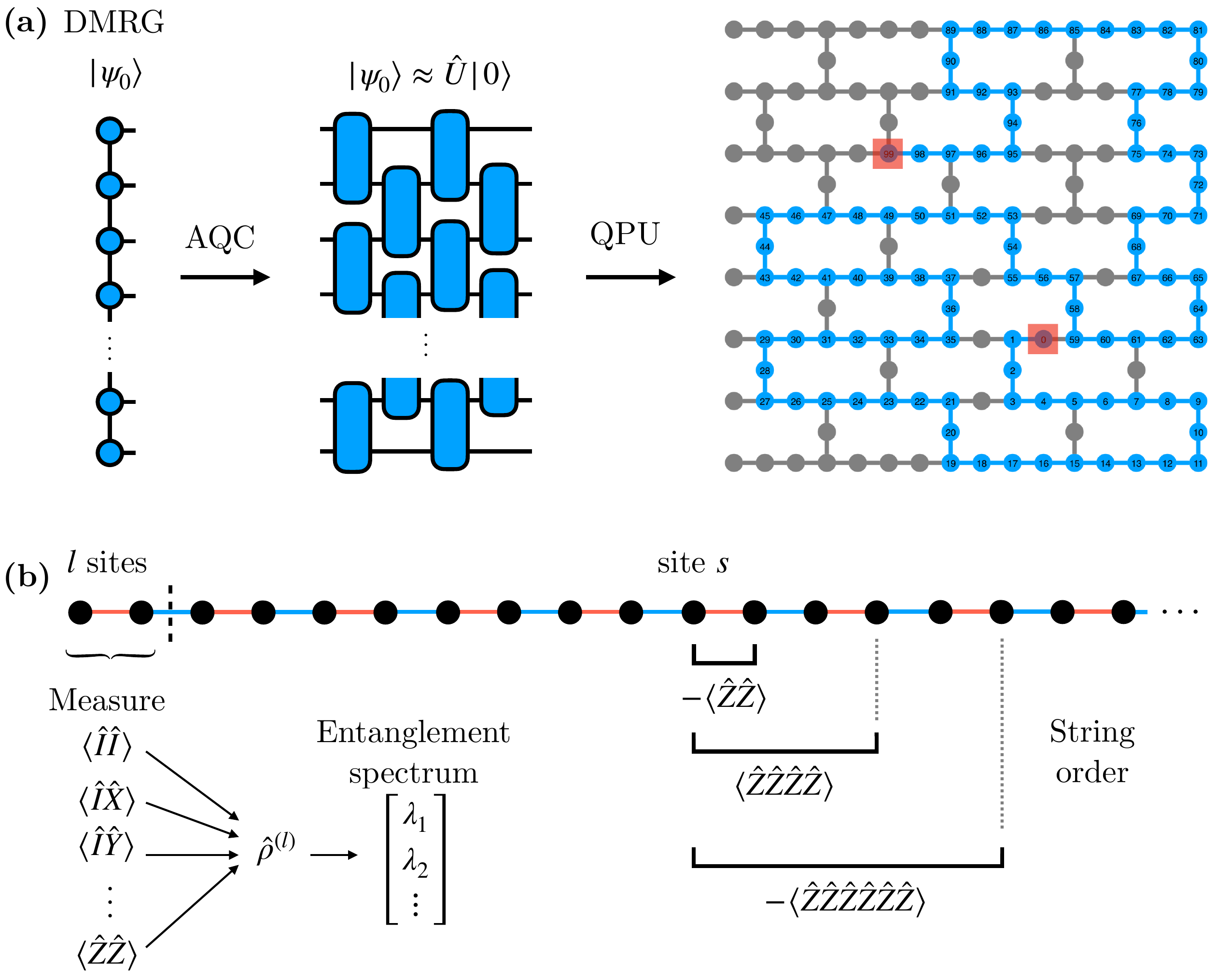}
    \caption{(a) A schematic of our workflow. We use \gls{dmrg} to find the ground state \gls{mps} of the Hamiltonian,  Eq.~\eqref{eq:bahc_hamiltonian}. Then we employ \gls{aqc}, specifically using the AQC-Tensor algorithm with a brickwork ansatz, to approximately map the target \gls{mps} to a shallow quantum circuit. Finally, we execute the circuit on quantum hardware. The coupling map shown is that of \texttt{ibm\_pittsburgh}, with the blue qubits and connections signifying those used for the experiment shown in Fig.~\ref{fig:3_string_order}, and the ends of the chain highlighted in red. (b) A sketch illustrating how the string order, Eq.~\eqref{eq:string_order}, is measured (right), and the tomography procedure for obtaining the entanglement spectrum (left). Here the $\lambda_i$'s are the sorted eigenvalues of the $l$-site reduced density matrix, $\hat{\rho}^{(l)}$, with $\lambda_1 \geq \lambda_2 \geq~...$}
    \label{fig:2_workflow}
\end{figure*}

with $J_i=J_0$ if $i$ is even, and $J_i=J_1$ otherwise. Here $\hat{X}$, $\hat{Y}$, and $\hat{Z}$ are the Pauli operators. The model is characterised by alternating nearest-neighbour magnetic coupling strengths $J_0$ and $J_1$ as depicted in Fig.~\ref{fig:1_model}a. The phase diagram~\cite{haghshenas2014symmetry} of this model is shown schematically in Fig.~\ref{fig:1_model}b, with the black crosses corresponding to the ground states considered in this work. There is a topologically-trivial ferromagnetic phase where both couplings are negative, as well as two phases exhibiting non-trivial \gls{spt} order, known as the odd-Haldane and even-Haldane phases. We consider four ground states in this work: three in the even-Haldane phase, with $J_0=1$ and $J_1=0.5,~-1,~-2$, and one in the odd-Haldane phase, with $J_0=0.5,~J_1=1$. We note that, since $J_0>0$ in the even-Haldane phase and $J_1>0$ in the odd-Haldane phase, and the ground state is unchanged by a global rescaling of the Hamiltonian, we can identify each ground state by the ratio of the couplings, and label them as $E_{J_1/J_0}$ in the even-Haldane phase and $O_{J_0/J_1}$ in the odd-Haldane phase. Using this notation, we refer to the ground states considered in this work as $O_\frac{1}{2}$, $E_\frac{1}{2}$, $E_{-1}$ and $E_{-2}$. Along the positive $J_0$ and $J_1$ axes, the ground states are simple products of singlet pairs, with free, uncoupled, edge spins in the odd-Haldane phase (see Fig.~\ref{fig:1_model}a). It is interesting to note that the $O_\frac{1}{2}$ ground state considered in this work is similar to the state experimentally realised in nanographene in Ref.~\cite{zhao2024tunable}, with $N=22$ sites and the parametrisation $J_0/J_1\approx0.605$. Thus, realising this model on quantum hardware is directly relevant to experimentally realisable materials.

Unlike the ferromagnetic phase, which is characterised by a local order parameter (magnetisation), the \gls{spt} phases are characterised by two non-local order parameters. In these phases, there is no conventional magnetic order: the single-site magnetisation, $\braket{\hat{Z}_i}$, is zero in the bulk, and standard two-point correlations, $\braket{\hat{Z}_i \hat{Z}_j}$, vanish exponentially with increasing separation. However, there is a form of hidden antiferromagnetic order, which is exposed by the non-local string order parameters. These are defined as the limit of expectation values of even-length Pauli strings~\cite{wang2013topological}:

\begin{equation}
    S^{\mathrm{E}/\mathrm{O}}= \lim_{l \rightarrow \infty} \left(- \Braket{\hat{Z}_s \mathrm{exp}\left( \sum_{i=s+1}^{s+l-2}\frac{i\pi}{2}\hat{Z}_i \right)\hat{Z}_{s+l-1} } \right),
    \label{eq:string_order}
\end{equation}

where $s$ denotes the left-most index of each string, and is even (odd) for the even (odd) string order parameter. $l$ corresponds to the string length, and is even for both parameters. Intuitively, the string order measures the correlations between two sites, with the addition of a non-local phase dependent on the sites in between. Whilst conventional correlations vanish exponentially, the even (odd) string order parameter converges to a non-zero value in the even (odd) Haldane phase. For ease of notation and nomenclature, we refer to each length-$l$ observable in the sequence in Eq.~\eqref{eq:string_order}:

\begin{equation}
    S^{\mathrm{E}/\mathrm{O}}_{l,s}= (-1)^{l / 2} \Braket{ \bigotimes_{i=s}^{s+l-1}\hat{Z}_i },
    \label{eq:finite_string_order}
\end{equation}

as the length-$l$ string order parameter. Since the length-$l$ string order rapidly tends toward its limiting value at a rate set by the correlation length (here only a few sites), it provides an excellent proxy for the true, infinite-length, string order. 

\begin{figure*}[ht]
    \centering
    \includegraphics[width=\linewidth]{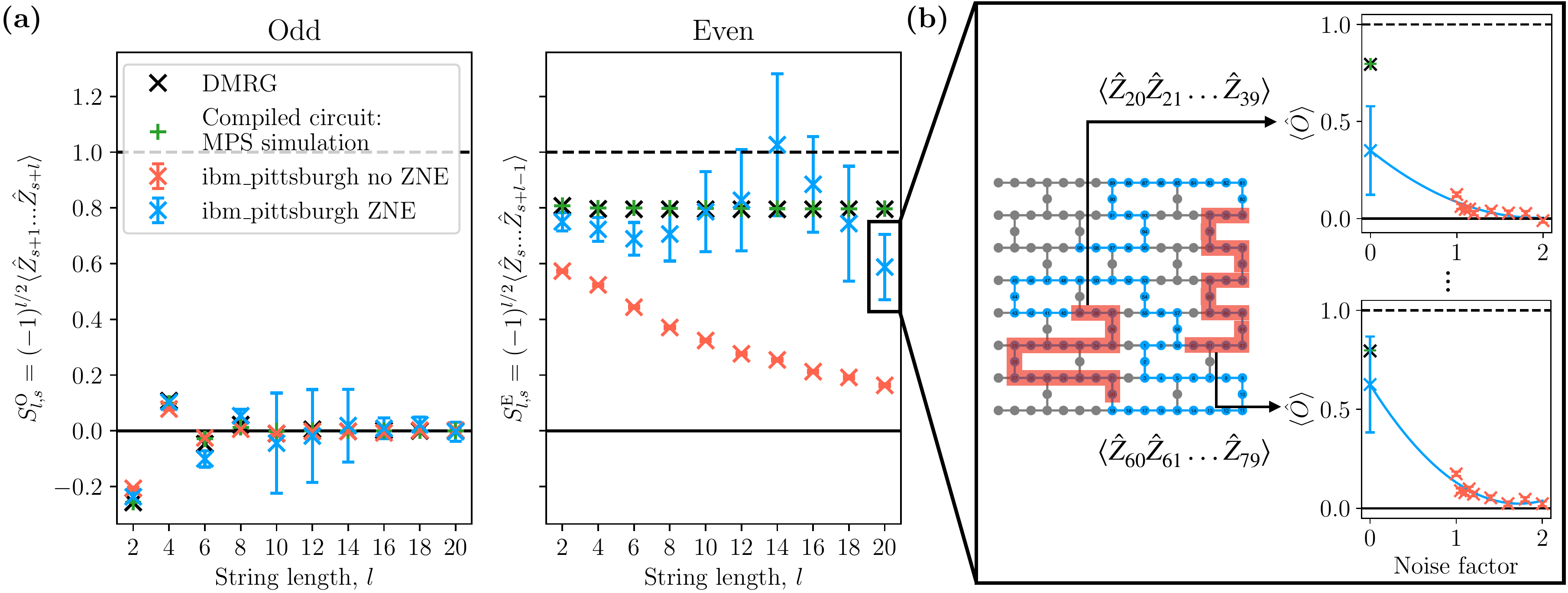}
    \caption{(a) Odd (left) and even (right) string order parameters as a function of string length for the $E_{-2}$ ground state, prepared and measured using \texttt{ibm\_pittsburgh}. Each data point is averaged over five sections of the chain, corresponding to $s=20,~30,~40,~50$, and 60. The blue (red) crosses show the expectation values with (without) \gls{zne}. The error bars represent the standard error on the mean (red) and the uncertainty of the fit evaluated at a noise factor of zero (blue), after propagation for the average value using the relation for a sum. The black and green crosses show the corresponding values obtained classically from \gls{dmrg} and from \gls{mps} simulation of the compiled circuit, respectively. The black dashed line represents the maximum possible value of 1. (b) A schematic of how the $l=20$ even string order parameter is obtained by averaging over multiple sections of the chain. (Left) \texttt{ibm\_pittsburgh} coupling map showing the qubits used in this experiment (blue) along with the footprints of the 20-local even string order observables (red) for $s=20$ and $60$. (Right) \gls{zne} curves for the two observables. The red crosses show the measured values as a function of noise factor, and the blue curve shows the extrapolation.}
    \label{fig:3_string_order}
\end{figure*}

In addition to the string order parameters, a stronger indicator for the presence of \gls{spt} order can be found in the entanglement spectrum of a state~\cite{pollmann2010entanglement}. The entanglement spectrum consists of the eigenvalues of the reduced density matrix of a bipartition of the chain, and contains information about how the two parts of the bipartition are entangled with each other. If the eigenvalues in the entanglement spectrum all have even degeneracy, then the entanglement across the bipartition cannot be adiabatically removed without either crossing a phase transition or breaking the symmetry of the system. On the other hand, if the eigenvalues all have odd degeneracy, the entanglement can be adiabatically removed. As detailed in Ref.~\cite{haghshenas2014symmetry}, the entanglement spectrum arising from a bipartition of the chain formed by cutting one of the $J_0$ ($J_1$) bonds exhibits eigenvalues with even (odd) degeneracy in the even-Haldane phase, signifying that the entanglement across $J_0$ bonds is protected by the symmetry, whereas the entanglement across the $J_1$ bonds is not. This is best understood by considering the case where the $J_1$ coupling is adiabatically weakened to zero, in which case the ground state becomes a product of maximally-entangled singlet pairs coupled by the $J_0$ bonds. In the odd-Haldane phase, the entanglement spectra exhibit a complementary pattern, where this time the $J_1$ ($J_0$) bonds exhibit even (odd) degeneracy. The alternating odd/even degeneracy in the entanglement spectrum is the most robust fingerprint of the \gls{spt} order. Whilst the above discussion strictly applies to infinite chains, in the finite chain case, a similar pattern should be observed when considering the entanglement spectra of $l$-site segments from the end of the chain, where the degeneracies should tend towards those of the infinite chain case, over a length scale comparable to the correlation length~\cite{smith2023deterministic}. 

Finally, since we consider the \gls{bahc} with $J_0$ terminations, the entanglement across the bonds at either end of the chain can be adiabatically removed in the odd-Haldane phase, resulting in a decoupled free spin at each end of the chain when $J_0=0$. In the case where $J_0 \neq 0$, we expect to see the free spin confined to the edge with a finite correlation length. In the even-Haldane phase, the entanglement across the $J_0$ bonds cannot be adiabatically removed without breaking the symmetry, and thus in this phase, with these terminations, there is not a free edge spin.

\section*{Ground state preparation on quantum hardware}

Figure~\ref{fig:2_workflow}a gives an overview of the routine used to prepare the ground states on quantum hardware. We first obtain the \gls{mps} representation of each of the four ground states using the \gls{dmrg} algorithm as implemented in the TeNPy library~\cite{tenpy2024}. The resulting \glspl{mps} have bond dimensions of between 19 and 40. See Methods Sec.~\ref{subsec:dmrg} for more details of the \gls{dmrg} procedure.

\begin{figure*}
    \centering
    \includegraphics[width=\linewidth]{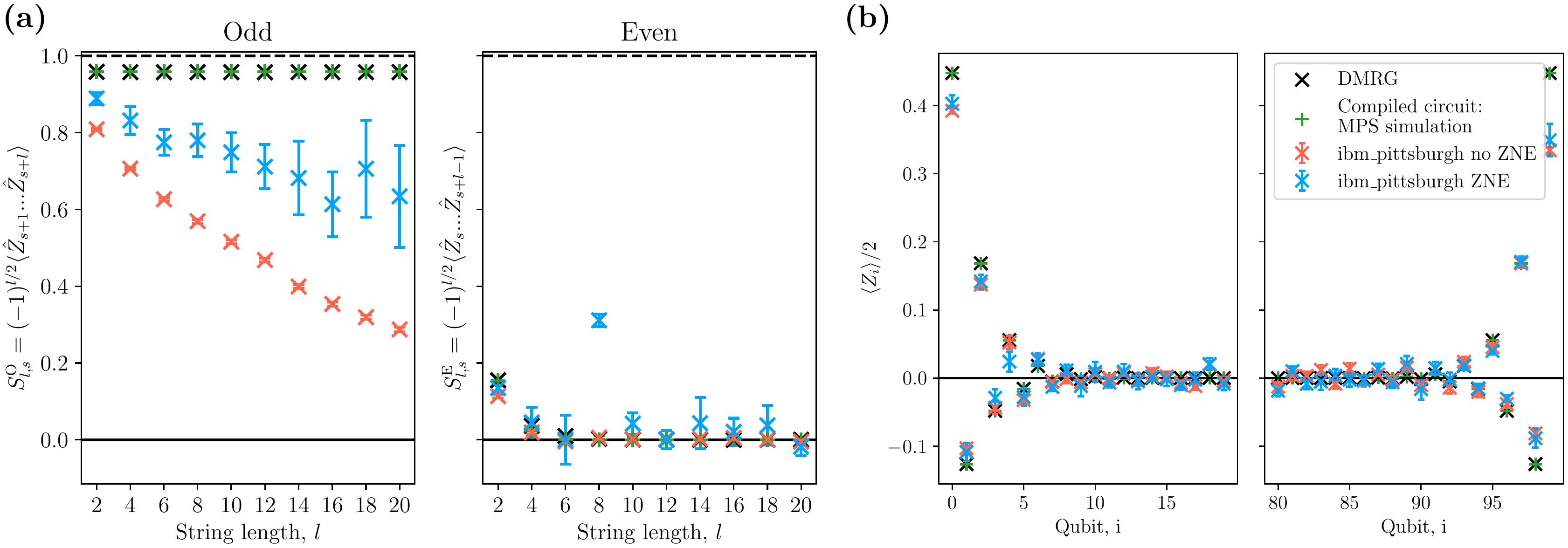}
    \caption{String order parameters and single-site magnetisation of the $O_\frac{1}{2}$ ground state, prepared and measured using \texttt{ibm\_pittsburgh}. (a) Odd (left) and even (right) string order parameters as a function of string length. Each data point is averaged over five sections of the chain, corresponding to $s=20,~30,~40,~50$, and 60. The black dashed line represents the maximum possible value of 1. (b) Single-site magnetisation, $\braket{\hat{Z}_i}/2$, of the twenty sites closest to each end of the chain. The fluctuations at the edges correspond to the exponentially-localised spin-1/2 edge states. In both panels, the blue (red) crosses show the expectation values with (without) \gls{zne}. The black and green crosses show the corresponding values obtained classically from \gls{dmrg} and from \gls{mps} simulation of the compiled circuit, respectively. The error bars represent the standard error on the mean (red) and the uncertainty of the fit evaluated at a noise factor of zero (blue). In panel (a), the uncertainties on the mean over the five sections are calculated using the relation for a sum.}
    \label{fig:4_edge_states}
\end{figure*}

Next, we compile quantum circuits to approximately prepare the \glspl{mps} using the AQC-Tensor algorithm~\cite{robertson2025approximate, qiskit-addon-aqc-tensor}. AQC-Tensor uses tensor networks to evaluate the fidelity $\lvert\braket{\psi|\hat{U}(\vec{\theta})|0}\rvert^2$ between a circuit ansatz, $\hat{U}(\vec{\theta})\ket{0}$, and a target state $\ket{\psi}$, and classically optimises the parameters, $\vec{\theta}$, to maximise the fidelity. To reduce computational overhead, we first compress the ground states whilst maintaining at least $99.9\%$ fidelity, resulting in \glspl{mps} with bond dimensions of between 5 and 8, and use these compressed \glspl{mps} as the target states for compilation, $\ket{\psi}$. However, we report all fidelities with respect to the uncompressed states. \gls{mps} simulation of 1D gapped ground states is known to be computationally efficient~\cite{vidal2004efficient} which has been leveraged previously to use AQC-Tensor to prepare ground states of such systems~\cite{jaderberg2025variational}. Due to the exponentially-decaying correlations~\cite{nachtergaele2006lieb} present in the ground states considered in this work, we use a brickwork circuit ansatz. Using this method, we obtain circuits which prepare the $O_\frac{1}{2}$, $E_\frac{1}{2}$, $E_{-1}$, and $E_{-2}$ ground states to fidelities of $98.9\%$, $98.9\%$, $99.0\%$ and $97.9\%$ with CNOT depths of 18, 21, 21, and 39, respectively. See Methods Sec.~\ref{subsec:aqc} for more details of the \gls{aqc} procedure, including the choice of ansatz and the initialisation method. Whilst standard methods for \gls{mps} preparation often use staircase circuits~\cite{schon2005sequential, lin2021real, ran2020encoding, schon2007sequential, rudolph2023decomposition}, we find that these methods were unable to reach comparable fidelities to \gls{aqc} without prohibitively high circuit depths, for the ground states considered in this work. We refer the reader to Supplementary Information~\ref{supp:comparison_to_other_techniques} for details of the fidelities and CNOT depths obtained by these methods.

\section*{Quantum hardware experiments}\label{sec:results}

In this section, we present experimental results for the measurement of the string order, edge magnetisation, and the entanglement spectrum, following the procedures sketched in Fig.~\ref{fig:2_workflow}b. See Methods Sec.~\ref{subsec:experimental_details} for details of the experimental procedures.

% Even Haldane string order.

Fig.~\ref{fig:3_string_order}a shows the length-$l$ string order parameters of Eq.~\eqref{eq:finite_string_order}, $S^{\mathrm{E}/\mathrm{O}}_{l,s}$, with length $l=2$ to $20$ for the $E_{-2}$ ground state, as measured on \texttt{ibm\_pittsburgh}~\cite{ibmq_systems}. Each data point is averaged over five sections of the chain, corresponding to different starting sites for the strings: $s=20,~30,~40,~50$, and 60. Fig.~\ref{fig:3_string_order}b illustrates the averaging procedure for the $l=20$ even string order parameter, along with the device footprint for the $S^\mathrm{E}_{l=20,s=20}$ and $S^\mathrm{E}_{l=20,s=60}$ observables. We exclude the first and last 20 sites so that the observables lie within the bulk, minimising edge effects. The red crosses show the expectation values with only the Pauli-twirling~\cite{wallman2016noise} and \gls{trex}~\cite{vandenberg2022model} error mitigation techniques applied, and the blue crosses show the expectation values with additional \gls{zne}~\cite{cai2021multi}. For details of the experimental implementation and error mitigation techniques, see Methods Sec.~\ref{subsubsec:string_order}. The black and green crosses show the expectation values calculated classically from \gls{dmrg} and from \gls{mps} simulation of the compiled circuit, respectively. Notably, the approximation error from the compilation is small enough that these values are almost indistinguishable.

\begin{figure*}[ht]
    \centering
    \includegraphics[width=\linewidth]{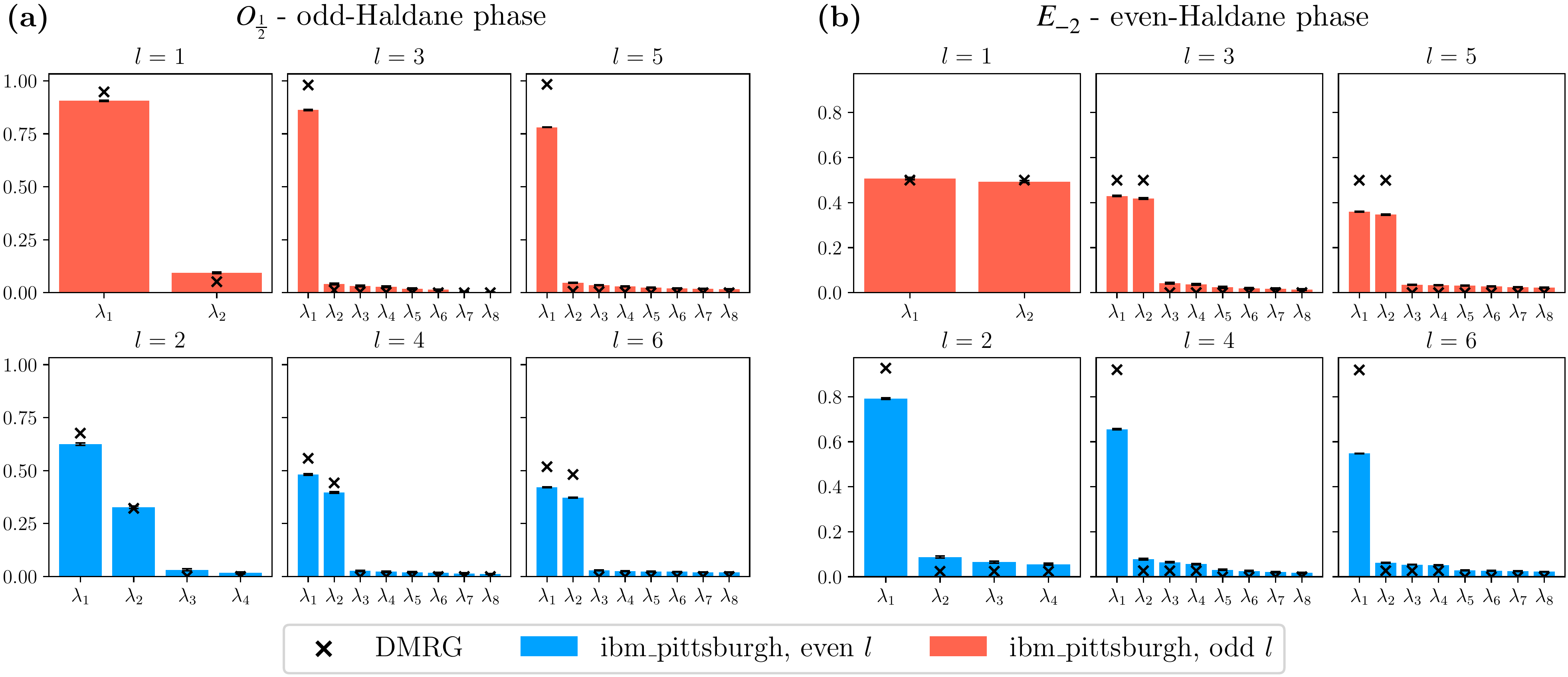}
    \caption{Entanglement spectra of the (a) $O_\frac{1}{2}$ and (b) $E_{-2}$ ground states for cuts of up to $l=6$ sites, executed on \texttt{ibm\_pittsburgh}. The red (blue) bars show the eigenvalues of the reduced density matrix in descending order,  $\lambda_1 \geq\lambda_2\geq\ldots$,  obtained by cutting $J_0$ ($J_1$) bonds. The error bars are obtained via a bootstrapping procedure and represent the standard deviation over 1000 samples of the Pauli-string expectation values. See Methods Sec.~\ref{subsubsec:entanglement_spectrum} for more details. The black crosses show the corresponding values obtained from \gls{dmrg}.}
    \label{fig:5_entanglement_spectrum}
\end{figure*}

We observe that $S_{l,s}^\mathrm{O}$ converges towards zero as the string length increases, consistent with the absence of odd string order in the even-Haldane phase. On the other hand, $S_{l,s}^\mathrm{E}$ remains non-zero even for strings of length $l=20$, indicating the presence of long-range \gls{spt} order. Since this is true even for the measured data before applying \gls{zne}, it implies that the \gls{spt} order is physically present in the state prepared on the quantum device, rather than an artifact of extrapolation. Results for the other two ground states in the even-Haldane phase, $E_\frac{1}{2}$ and $E_{-1}$, are shown in Supplementary Information~\ref{supp:further_results}, in Fig.~\ref{fig:J0_1.0_J1_0.5_string_order} and Fig.~\ref{fig:J0_1.0_J1_-1.0_string_order}, respectively. These results are consistent with those of the $E_{-2}$ ground state, shown here.

% Odd Haldane string order and edge magnetisation.

Fig.~\ref{fig:4_edge_states}a shows the corresponding $S^{\mathrm{E}/\mathrm{O}}_{l,s}$ values for the $O_\frac{1}{2}$ ground state. The meaning of all elements of the plot is the same as in Fig.~\ref{fig:3_string_order}. In this case we observe complementary behaviour to that in the even-Haldane phase, with $S_{l,s}^\mathrm{E}$ converging towards zero as the string length increases, and $S_{l,s}^\mathrm{O}$ remaining non-zero up to length $l=20$. This is consistent with the presence of odd string order and the absence of even string order in the odd-Haldane phase.

Following on from the string order results, Fig.~\ref{fig:4_edge_states}b shows the single-site magnetisation, $\braket{\hat{Z}_i}/2$, of the $O_\frac{1}{2}$ ground state, for the twenty sites closest to the left end of the chain. These expectation values were measured during the same experiment as the string order, using the same error mitigation techniques, and all elements of the plot have the same meaning. In this case, we use linear fits for \gls{zne}, since the expectation values were fairly constant across the range of noise factors. We observe that the expectation values obtained from the hardware agree well with those obtained classically, and exhibit antiferromagnetic ordering close to the edges, with the magnetisation decaying to zero in the bulk.

As discussed previously, the odd-Haldane phase exhibits a symmetry protected, exponentially-localised, spin-1/2 edge state, which we observe as a net magnetisation of 1/2 close to each end of the chain. By grouping lattice sites into two-site unit cells, we fit the magnetisation of each unit cell, $S^z_1=(\braket{\hat{Z}_{2i}}+\braket{\hat{Z}_{2i+1}})/2$, from the left end of the chain, to the functional form: $S^z_1\propto e^{-x/\xi_1}$. From this, we extract a characteristic correlation length, $\xi=2\xi_1$, of $\xi=1.97 \pm 0.03$ (\gls{dmrg} and \gls{aqc}), $1.76\pm0.20$ (hardware, no \gls{zne}), and $1.81\pm0.21$ (hardware, linear \gls{zne}). See Fig.~\ref{fig:edge_decay_fits} in Supplementary Information~\ref{supp:edge_decay} for the fits. In principle, the correlation length can be used to estimate the energy gap, $\Delta$, using the relation $\Delta\sim v/\xi$~\cite{hastings2006spectral}, provided a suitable estimate for the excitation group velocity, $v$, can be obtained. Here we simply note that the length scale over which the string order is measured is an order of magnitude larger than the correlation length.

% Entanglement spectrum.

Finally, we obtain the entanglement spectra of the $O_\frac{1}{2}$ and $E_{-2}$ ground states via state tomography. We construct the reduced density matrices of $l$-site segments from the end of the chain for $l=1$ to 6, by measuring the expectation values of all $l$-site Pauli strings and combining the values using Eq.~\eqref{eq:tomography}, as illustrated in Figure~\ref{fig:2_workflow}b. Our results are shown in Fig.~\ref{fig:5_entanglement_spectrum}. The red (blue) bars represent the eight largest eigenvalues in the spectra obtained by cutting $J_0$ ($J_1$) bonds, as measured on \texttt{ibm\_pittsburgh}. Here we use the Pauli-twirling and \gls{trex} error mitigation techniques, but not \gls{zne} due to the increased overhead arising from the fact that not all of the observables commute. The black error bars in the middle of each bar are obtained using a bootstrapping procedure, and show the standard deviation of the values over 1000 samples of the Pauli-string expectation values. See Methods~\ref{subsubsec:entanglement_spectrum} for details of the experimental procedure. The black crosses denote the theoretical expectation, as obtained by \gls{dmrg}. We do not show the values obtained via \gls{mps} simulation of the \gls{aqc} circuits, since at this scale they are indistinguishable from those from \gls{dmrg}. See Supplementary Information~\ref{supp:aqc_on_entanglement} for a detailed look at the effect of the \gls{aqc} approximation error.

For the $E_{-2}$ ground state (Fig.~\ref{fig:5_entanglement_spectrum}b), we observe an alternating pattern where the entanglement spectra arising from cuts to the $J_0$ bonds exhibit two dominant eigenvalues, whereas the spectra from cuts to the $J_1$ bonds exhibit only a single dominant eigenvalue. This observation is qualitatively consistent with the theoretical expectation that the entanglement spectra across $J_0$ bonds should exhibit even degeneracy, arising from the fact that the entanglement across those bonds cannot be adiabatically removed without crossing a phase boundary, or breaking the symmetry. Whilst we cannot quantitatively argue that the eigenvalues in the spectra are degenerate, we note that this is to be expected, since even a slight error in the reduced density matrix would lead to the eigenvalues not being perfectly degenerate. Results for the other even-Haldane phase ground states, $E_\frac{1}{2}$ and $E_{-1}$, are shown in Supplementary Information~\ref{supp:further_results}, in Fig.~\ref{fig:J0_1.0_J1_0.5_entanglement_spectrum} and Fig.~\ref{fig:J0_1.0_J1_-1.0_entanglement_spectrum}, respectively, and are consistent with the $E_{-2}$ results. Looking at the $O_\frac{1}{2}$ ground state (Fig.~\ref{fig:5_entanglement_spectrum}a), we observe complementary behaviour to the even-Haldane phase, where in this phase, the spectra arising from cuts to the $J_1$ bonds exhibit two dominant eigenvalues, and those from cuts to $J_0$ bonds exhibit only a single dominant eigenvalue. In contrast to the even-Haldane phase, in this case the degeneracies of the eigenvalues are only reached with increasing $l$. This is expected due to the influence of the confined edge mode. The suggestion of even degeneracy in the entanglement spectrum across $J_0$ ($J_1$) bonds in the even (odd) Haldane phases is a strong indication of the signature symmetry-protection of the entanglement across these bonds.

It should be noted that the even degeneracy of the two eigenvalues in the $l=1$ spectrum for the $E_{-2}$ ground state would also be observed under complete decoherence, where the reduced density matrix of the first site would be described by the maximally mixed state. However, the results from the other spectra indicate that this is not the case, due to the alternating pattern of odd/even degeneracy within each phase, and the complementary patterns observed when comparing the two phases. Both of these observations combine to give strong evidence of the successful preparation of states with \gls{spt} order on the quantum hardware. 

\section*{Conclusion}\label{sec:conclusion}

% Core result
In this work, we have demonstrated the preparation of 100-site ground states in the \gls{spt} phases of the \gls{bahc} on an IBM superconducting quantum processor. This was enabled by the combination of \gls{dmrg} and tensor network \gls{aqc}, producing high-fidelity ($97.9$-$99.0\%$), shallow quantum circuits (18-39 CNOT depth). For ground states spanning both the even-Haldane and odd-Haldane \gls{spt} phases, we verified the prepared states using multiple characteristic signatures of \gls{spt} order. We have experimentally observed significant, non-zero string order, qualitative agreement with the expected degeneracies in the entanglement spectrum and, in the odd-Haldane phase, the presence of exponentially-localised, symmetry protected edge states.

% Scientific significance
Importantly, the ability to prepare large, entangled ground states to high fidelity on quantum hardware is an essential first step in the study of more complex many-body phenomena. In particular, access to high-quality \gls{spt} ground states enables investigations of quench dynamics and other non-equilibrium processes in regimes which become increasingly difficult to simulate classically due to entanglement growth. The successful preparation of \gls{spt} states also has applications in measurement-based quantum computation~\cite{else2012symmetry, daniel2021quantum}.

% Limitations/scope
The method presented in this work is well suited to the preparation of ground states of 1D gapped Hamiltonians with short-range correlations, as demonstrated with the bond-alternating Heisenberg chain. However, for states with long-range entanglement or large correlation lengths, for example, excited states or ground states of gapless Hamiltonians, or in cases where a good \gls{aqc} initialisation is not known, the method will struggle. Indeed, as detailed in Methods Sec.~\ref{subsec:aqc}, the $E_{-2}$ ground state was much harder to compile than the other states considered, due to its comparatively large correlation length. We note that this is not a failure of \gls{aqc}, but a theoretically expected trend, as Ref.~\cite{malz2024preparation} showed that the circuit depth required to faithfully prepare translationally invariant, normal \glspl{mps} scales logarithmically in system size and linearly in correlation length.  In such cases, alternative ans\"atze which are better suited for capturing long-range correlations may be required. It is also interesting to consider if symmetries of the system can be exploited to reduce the number of independent variational parameters in the ansatz, something we did not utilise in this work. Finally, a potentially fruitful direction is to generalise AQC to incorporate adaptive circuit elements, namely mid-circuit measurements and classical feedforward, which are known to be powerful tools in \gls{mps} preparation~\cite{smith2024constant}.

% Forward-looking
Looking forward, our results lay the foundation for several promising directions. One opportunity in the immediate future is the use of digital quantum computers to verify experimental results in \gls{spt} physics, such as the recent realisation of the \gls{bahc} in nanographene for up to 40 sites~\cite{zhao2024tunable}. Another direction is to extend the simulations performed in this work, for example, by probing the excitation spectrum of the model using local quench spectroscopy~\cite{villa2020local} or by preparing ground states of more complex models such as those with defects or interfaces between different regions. Finally, investigating how our method can be extended to higher-dimensional geometries, where classical simulation methods particularly struggle, is of high priority. Overall, our work represents an important first step toward performing classically challenging many-body simulations of out-of-equilibrium phenomena in \gls{spt} systems on quantum hardware.

%TC:ignore
\section*{Methods}\label{sec:methods}

\subsection{DMRG}\label{subsec:dmrg}

For the ground states considered in this work, we obtain their \gls{mps} representations using the \gls{dmrg} algorithm, as implemented in the TeNPy library~\cite{tenpy2024}. For the \gls{dmrg} procedure, we used at most 20 sweeps (\texttt{max\_sweeps}=$20$), the default mixer (\texttt{mixer}=\texttt{True}), and \texttt{combine}=\texttt{True}, to reduce computational overhead. When performing the singular value decomposition during \gls{dmrg}, we we keep up to 100 of the largest Schmidt values (\texttt{chi\_max}=$100$), discard any below $10^{-10}$ (\texttt{svd\_min}=$10^{-10}$), and discard the smallest values such that the sum of their squares is less than $10^{-12}$ (\texttt{trunc\_cut}=$10^{-6}$). We refer the reader to the TeNPy documentation for \texttt{DMRGEngine} and \texttt{truncate} for the precise meanings of these configuration options. The bond dimensions and energies of the four ground states considered in this work are shown in the top half of Table~\ref{tab:bahc_compiling_results}.

We note that the ground states in the even-Haldane phase are unique, whereas there is a four-fold degeneracy in the odd-Haldane phase. This is easiest understood by considering the $J_0=0$, $J_1>0$ axis, where the ground state consists of 49 singlet-pairs in the bulk, plus two free spin-1/2s at the edges. In this work, we take the $+1$ magnetisation ground state, corresponding to the edge spins being in the $\uparrow\uparrow$ configuration.

\begin{table}
    \centering
    \begin{tabular}{|c|c|c|c|c|}
        \hline
        State label & $O_\frac{1}{2}$ & $E_\frac{1}{2}$ & $E_{-1}$ & $E_{-2}$ \\
        \hline
        $J_0$ & 0.5 & 1 & 1 & 1 \\
        $J_1$ & 1 & 0.5 & -1 & -2 \\
        $\chi$ & 22 & 19 & 26 & 40 \\ 
        $E$ & -38.166 & -38.819 & -41.150 & -49.329\\
        $\chi$ (after compression) & 5 & 5 & 8 & 8 \\
        $E$ (after compression) & -38.164 & -38.817 & -41.150 & -49.329\\
        \hline
        Compiled fidelity & 0.989 & 0.989 & 0.990 & 0.979 \\
        Compiled energy & -38.142 & -38.794 & -41.121 & -49.268 \\
        brickwork layers, $L$ & 3 & 3.5 & 3.5 & 6.5 \\
        CNOT depth & 18 & 21 & 21 & 39 \\
        CNOT count & 891 & 1041 & 1041 & 1932 \\
        \hline
    \end{tabular}
    \caption{(Top) Bond dimensions, $\chi$, and energies, $E$, of the four ground states considered in this work. We also report the bond dimensions and energies after compressing to the lowest bond dimension maintaining at least $99.9\%$ fidelity. (Bottom) Compilation results. We report the compiled circuit with the highest fidelity for each ground state. If multiple circuits reach the same fidelity, we report the one with the lowest depth. Compilation is performed with the compressed states as the targets, but all fidelities are reported with respect to the original, uncompressed states obtained using \gls{dmrg}.}
    \label{tab:bahc_compiling_results}
\end{table}

\subsection{Approximate Quantum Compilation}\label{subsec:aqc}

We use \gls{aqc} to prepare the ground state \glspl{mps} as quantum circuits. The goal of \gls{aqc} is to optimise a variational circuit ansatz to maximise the fidelity with a target state of interest. More precisely, given a target state $\ket{\psi}$, and a variational circuit ansatz $\hat{U}(\vec{\theta})$, \gls{aqc} seeks to minimise the cost function:

\begin{equation}
    C=1-\lvert\braket{\psi|\hat{U}(\vec{\theta})|0} \rvert^2.
    \label{eq:global_cost}
\end{equation}

For the target states, $\ket{\psi}$, we use compressed versions of the \glspl{mps} obtained by \gls{dmrg} (see Sec.~\ref{subsec:dmrg}) maintaining at least $99.9\%$ fidelity. We perform the compression using the \texttt{variational} method in TeNPy, with between 10 and 50 sweeps. This incurs a small error in the cost function, Eq.\eqref{eq:global_cost}, compared to using the converged ground state, however the lower bond dimension reduces computational cost of evaluating the cost function and its gradients, which we deem to be a worthwhile compromise. Furthermore, the infidelity of the compressed \glspl{mps} ($\leq0.1\%)$ is chosen to be an order of magnitude lower than the desired infidelity for \gls{aqc} ($1\%$). We stress that, whilst we use compressed \glspl{mps} as the target states for \gls{aqc}, we report all fidelities with respect to the original, uncompressed ground states.

To perform \gls{aqc}, we use the AQC-Tensor add-on for Qiskit~\cite{qiskit-addon-aqc-tensor, robertson2025approximate}, which is an efficient implementation, using tensor networks to evaluate the cost function, Eq.~\eqref{eq:global_cost}, and its gradients.

An important factor for \gls{aqc} is the choice of ansatz. In this work, we use the $L$-layer nearest-neighbour brickwork ansatz, for its shallow depth and ability to capture short-range correlations. Defining a single brickwork layer as a set of two-qubit unitaries acting on all pairs of qubits $(i,~i+1)$ with even $i$, followed by an analogous set with odd $i$, the $L$-layer brickwork ansatz is able to exhibit non-zero correlations, $C_{zz}(i,~j)=\braket{\hat{Z}_i\hat{Z}_j}-\braket{\hat{Z}_i}\braket{\hat{Z}_j}$, for sites separated by up to $|i-j|=4L-1$. Since the Hamiltonian studied in this work is gapped, its ground states exhibit exponentially decaying correlations~\cite{nachtergaele2006lieb}, suggesting that they may be well-suited to the brickwork ansatz. It should be noted that the brickwork ansatz may not be well suited for ground states of gapless Hamiltonians, since they often (but not always) exhibit long-range, power-law decaying correlations~\cite{nachtergaele2006lieb}.

Each two-qubit unitary in the brickwork ansatz is a universal $\text{SU}(4)$ unitary. When preparing the circuit ansatz for AQC-Tensor, the following pre-processing steps are applied to the brickwork ansatz: 1. each two-qubit unitary is decomposed using a Cartan decomposition with 15 variational parameters~\cite{vatan2004optimal}, 2. any single-qubit unitaries are similarly decomposed into a sequence of $R_z R_x R_z$ gates, 3. the total number of variational parameters is reduced by combining sequences of single-qubit or two-qubit unitaries into corresponding universal $\text{SU}(4)$ or $\text{SU}(2)$ unitaries, and finally 4. initial $R_z$ rotations acting on the input state, $\ket{0}^{\otimes N}$ are removed.

We choose initial ansatz parameters such that the initial state, $\ket{\psi_0}=\hat{U}(\vec{\theta}_0)\ket{0}^{\otimes N}$, corresponds to the ground state of the model when one of the coupling strengths is zero. This corresponds to a product of singlet states: $\ket{\psi_{\text{even}}}=\left(\left(\ket{01}-\ket{10}\right)/\sqrt{2}\right)^{\otimes N//2}$ for the even-Haldane phase, and a product of singlet states plus two $\uparrow$-configuration edge spins: $\ket{\psi_{\text{odd}}}=\ket{0}\otimes\left(\left(\ket{01}-\ket{10}\right)/\sqrt{2}\right)^{\otimes N//2-1}\otimes\ket{0}$ for the odd-Haldane phase. These states can be prepared with a depth-1 circuit, which we append as an extra half brickwork layer to the ansatz for the even-Haldane phase, and incorporate into the final half-layer of the ansatz for the odd-Haldane phase. This initialisation resulted in initial fidelities of $40\%$, $46\%$, $23\%$, and $2\%$ for the $O_\frac{1}{2}$, $E_\frac{1}{2}$, $E_{-1}$, and $E_{-2}$ ground states, respectively. We note that, due to the singlet-like nature of the ground states, a product state initialisation is not viable. Indeed, we find that such an initialisation yields fidelities $\mathcal{O}(10^{-10})$ or lower for the ground states considered in this work, and hence was not used.

The bottom half of Table~\ref{tab:bahc_compiling_results} gives details of the compilation results for the four ground states. We executed multiple compilation runs in parallel, with $L=2.5$--3.5, 2.5--5.5, and 3.5--6.5 (integer increments) for each of the even-Haldane ground state respectively. For the $O_\frac{1}{2}$ ground state, we used $L=3$, since this was sufficient for the corresponding $E_\frac{1}{2}$ ground state. Of these parallel runs, we take the one which reached the highest fidelity, and if multiple executions reach the same fidelity, we take the one with the lowest depth. Using the \texttt{adam} optimiser~\cite{kingma2017adam}, the \gls{aqc} procedure was able to reach $98.9\%$, $98.9\%$, and $99.0\%$ fidelity for the $O_\frac{1}{2}$, $E_\frac{1}{2}$, and $E_{-1}$ ground states in a few hours using 3, 3.5, and 3.5 brickwork layers, respectively. In these cases, compilation exited when the target fidelity of $99\%$, with respect to the compressed target \gls{mps}, was reached. For the remaining $E_{-2}$ ground state, compilation managed to reach $97.9\%$ fidelity over a 7-day wall-time using 6.5 brickwork layers. We attribute the difficulty of compiling the latter state to the order-of-magnitude lower initial fidelity, and the increased computational overhead of evaluating the cost function and its gradients for the deeper ansatz and higher number of variational parameters.

\subsection{Experimental Details}\label{subsec:experimental_details}

For our experimental results, we execute the compiled circuits on quantum hardware and measure the string order, entanglement spectrum, and edge state magnetisation (odd-Haldane phase only). Here we provide details of the experimental setup. 

We transpile the compiled circuits to the device using the Qiskit transpiler~\cite{qiskit2024} using the default settings, and optimisation level 3. When choosing a mapping from virtual qubits to physical qubits on the device, the transpiler makes use of the latest calibration data, and attempts to avoid particularly error-prone qubits and connections.

\subsubsection{String order}\label{subsubsec:string_order}

The length-$l$ string order parameters of Eq.~\eqref{eq:finite_string_order}, $S^{\mathrm{E}/\mathrm{O}}_{l,s}$,correspond to expectation values of Pauli-Z strings of even-length, and left-most index $s$ (even for the even parameters and odd for the odd parameters). For example, $S^\mathrm{E}_{6,~20}=-\braket{\hat{Z}_{20}\hat{Z}_{21}\hat{Z}_{22}\hat{Z}_{23}\hat{Z}_{24}\hat{Z}_{25}}$. The string order parameters, Eq.~\eqref{eq:string_order}, correspond to the $l\rightarrow \infty$ limit.

We measure the length-$l$ string order parameters for $l=2,~4,~...,~20$ over five sections of the chain corresponding to $s=20,~30,~40,~50$, and $60$, avoiding 20 sites from either end of the chain to avoid edge effects. We execute the \gls{aqc} compiled circuits on the \texttt{ibm\_pittsburgh} quantum computer and measure the expectation values using the Qiskit~\cite{qiskit2024} \texttt{Estimator} primitive. For each observable, we use $10\,000$ shots, \gls{trex}~\cite{vandenberg2022model}, Pauli-twirling~\cite{wallman2016noise} with 100 randomisations, and \gls{zne}~\cite{cai2021multi}. For the \gls{zne} procedure, we use the noise factors: $1,~1.05,~1.1,~1.15,~1.2,~1.4,~1.6,~1.8$, and $2.0$ and we use the best-fit extrapolator, from the choice of exponential, quadratic, and linear. After obtaining the extrapolated values of $S^\mathrm{E}_{l,s}$ and $S^\mathrm{O}_{l,s}$, we average the values for each $l$ over the values of $s$, and propagate the corresponding uncertainties using the relation for a sum. We note that we use the same experimental setup when measuring the single-qubit magnetisations $\braket{\hat{Z}_i}$, but in that case we do not average over multiple windows, and we use homogeneous linear extrapolation.

Our results for the $E_{-2}$ and $O_\frac{1}{2}$ ground state experiments are shown in Fig.~\ref{fig:3_string_order} and Fig.~\ref{fig:4_edge_states}, and used 1m 35s and 1m 29s of QPU time, respectively.

Immediately before our circuit is executed on the quantum device, we execute a circuit with the same two-qubit gate structure, but the action of which, in the absence of noise, corresponds to the identity, and perform the \gls{zne} procedure on individual Pauli-Z observables. With the knowledge that $\braket{\hat{Z_i}}=1$ in the absence of noise, we can validate our choice of noise factors, and check for any particularly error-prone qubits. This is similar to the calibration procedure discussed in~\cite{majumdar2023best}. For more details on this validation technique, we refer the reader to Supplementary Information~\ref{supp:identity_zne}.

\subsubsection{Entanglement spectrum}\label{subsubsec:entanglement_spectrum}

We use tomography to measure the reduced density matrices of $l$-site segments, using the equation:

\begin{equation}
    \hat{\rho}^{(l)}=\frac{1}{2^l}\sum_{k_1,~...,~k_l}\braket{ \hat{\sigma}_{k_1}...\hat{\sigma}_{k_l}} \hat{\sigma}_{k_1}...\hat{\sigma}_{k_l},
    \label{eq:tomography}
\end{equation}

with each $\hat{\sigma}_{k_i}\in \{\hat{I},~\hat{X},~\hat{Y},~\hat{Z}\}$~\cite{Nielsen_Chuang_2010}. The number of required observables scales as $4^l$ which, even after grouping into commuting groups (206 as opposed to 4096 for $l=6$), incurs a significant resource overhead to measure. For this reason, we do not use \gls{zne} for this experiment. Whilst more efficient methods exist for performing state tomography~\cite{wang2025direct}, we use this method for simplicity.

Using the Qiskit~\cite{qiskit2024} \texttt{Estimator} primitive, we measure the expectation values of all length-$l$ Pauli strings on the \texttt{ibm\_pittsburgh} device. For each observable, we use $10\,000$ shots, \gls{trex}~\cite{vandenberg2022model}, and Pauli-twirling~\cite{wallman2016noise} with 100 randomisations. It should be noted that we do not need to repeat the experiment for all values of $l$, since all Pauli strings of length $a<l$ are contained in the set of Pauli strings of length $l$.

From the measured expectation values, we construct the reduced density matrix, $\hat{\rho}^{(l)}$, and compute its eigenvalues $\{\lambda_i\}$, with $\lambda_1\geq\lambda_2\geq~...$. To estimate the uncertainties of the eigenvalues we use a bootstrapping procedure. We generate $k=1\,000$ samples of $\hat{\rho}^{(l)}$, denoted as $\hat{\rho}^{(l)}_j$ for $j \in \{1,~...,~k\}$ by sampling $k$ values of each Pauli string expectation value $\braket{\hat{\sigma}_{k_1}...\hat{\sigma}_{k_l}}$, from a normal distribution with the same mean and standard deviation as measured from the quantum device. Then we transform the sampled matrix $\hat{\rho}^{(l)}_j$ into the eigenbasis of the average matrix $\hat{\rho}^{(l)}$, and take the diagonal entries $\{\lambda^j_i\}$. We then take the mean and standard deviation of each value, $\lambda_i^j$ over the $k$ samples ($j=1,~...,~k$).

Our results for the $O_\frac{1}{2}$ and $E_{-2}$ ground state experiments, using $l=6$, are shown in Fig.~\ref{fig:5_entanglement_spectrum}a and b, requiring 44m 46s and 48m 18s of QPU time, respectively. \\

\begin{acknowledgments}

This work was supported by the Hartree National Centre for Digital Innovation, a collaboration between the Science and Technology Facilities Council and IBM. L.P.L. acknowledges support from the Engineering and Physical Sciences Research Council (Grant No. EP/Z53318X/1).

We would like to thank Stefano Mensa and Christa Zoufal for their advice and technical discussions. \\

\end{acknowledgments}

\section*{Contributions}

BJ, JC, and GP conceptualised the work, GP and BJ conducted the experiments, GP, BJ, and JC wrote first draft of the paper. All authors contributed technical expertise, participated in technical discussions, and reviewed and edited the paper.

\bibliography{bibliography.bib}

@misc{jaderberg2025variational,
      title={Variational preparation of normal matrix product states on quantum computers}, 
      author={Ben Jaderberg and George Pennington and Kate V. Marshall and Lewis W. Anderson and Abhishek Agarwal and Lachlan P. Lindoy and Ivan Rungger and Stefano Mensa and Jason Crain},
      year={2025},
      eprint={2503.09683},
      archivePrefix={arXiv},
      primaryClass={quant-ph},
      url={https://arxiv.org/abs/2503.09683}, 
}

@article{smith2023deterministic,
  title = {Deterministic Constant-Depth Preparation of the {AKLT} State on a Quantum Processor Using Fusion Measurements},
  author = {Smith, Kevin C. and Crane, Eleanor and Wiebe, Nathan and Girvin, S.M.},
  journal = {PRX Quantum},
  volume = {4},
  issue = {2},
  pages = {020315},
  numpages = {24},
  year = {2023},
  month = {Apr},
  publisher = {American Physical Society},
  doi = {10.1103/PRXQuantum.4.020315},
  url = {https://link.aps.org/doi/10.1103/PRXQuantum.4.020315}
}

@article{haghshenas2014symmetry,
doi = {10.1088/0953-8984/26/45/456001},
url = {https://doi.org/10.1088/0953-8984/26/45/456001},
year = {2014},
month = {oct},
publisher = {IOP Publishing},
volume = {26},
number = {45},
pages = {456001},
author = {Haghshenas, R and Langari, A and Rezakhani, A T},
title = {Symmetry fractionalization: symmetry-protected topological phases of the bond-alternating spin-1/2 Heisenberg chain},
journal = {Journal of Physics: Condensed Matter}
}

@article{schon2005sequential,
  title={Sequential generation of entangled multiqubit states},
  author={Sch{\"o}n, Christian and Solano, Enrique and Verstraete, Frank and Cirac, J Ignacio and Wolf, Michael M},
  journal={Physical review letters},
  volume={95},
  number={11},
  pages={110503},
  year={2005},
  publisher={APS}
}

@article{schon2007sequential,
  title = {Sequential generation of matrix-product states in cavity QED},
  author = {Sch\"on, C. and Hammerer, K. and Wolf, M. M. and Cirac, J. I. and Solano, E.},
  journal = {Phys. Rev. A},
  volume = {75},
  issue = {3},
  pages = {032311},
  numpages = {10},
  year = {2007},
  month = {Mar},
  publisher = {American Physical Society},
  doi = {10.1103/PhysRevA.75.032311},
  url = {https://link.aps.org/doi/10.1103/PhysRevA.75.032311}
}

@article{ran2020encoding,
  title={Encoding of matrix product states into quantum circuits of one-and two-qubit gates},
  author={Ran, Shi-Ju},
  journal={Physical Review A},
  volume={101},
  number={3},
  pages={032310},
  year={2020},
  publisher={APS}
}

@article{lin2021real,
  title = {Real- and Imaginary-Time Evolution with Compressed Quantum Circuits},
  author = {Lin, Sheng-Hsuan and Dilip, Rohit and Green, Andrew G. and Smith, Adam and Pollmann, Frank},
  journal = {PRX Quantum},
  volume = {2},
  issue = {1},
  pages = {010342},
  numpages = {15},
  year = {2021},
  month = {Mar},
  publisher = {American Physical Society},
  doi = {10.1103/PRXQuantum.2.010342},
  url = {https://link.aps.org/doi/10.1103/PRXQuantum.2.010342}
}

@article{wang2013topological,
  title = {Topological quantum phase transition in bond-alternating spin-$\frac{1}{2}$ Heisenberg chains},
  author = {Wang, Hai Tao and Li, Bo and Cho, Sam Young},
  journal = {Phys. Rev. B},
  volume = {87},
  issue = {5},
  pages = {054402},
  numpages = {6},
  year = {2013},
  month = {Feb},
  publisher = {American Physical Society},
  doi = {10.1103/PhysRevB.87.054402},
  url = {https://link.aps.org/doi/10.1103/PhysRevB.87.054402}
}

@article{robertson2025approximate,
       author = {{Robertson}, Niall F. and {Akhriev}, Albert and {Vala}, Jiri and {Zhuk}, Sergiy},
        title = "{Approximate Quantum Compiling for Quantum Simulation: A Tensor Network Based Approach}",
      journal = {{ACM Transactions on Quantum Computing}},
         year = {2025},
   issue_date = {September 2025},
    publisher = {Association for Computing Machinery},
      address = {New York, NY, USA},
       volume = {6},
       number = {3},
          url = {https://doi.org/10.1145/3731251},
          doi = {10.1145/3731251},
        month = may,
    articleno = {20},
     numpages = {15}
}

@misc{qiskit-addon-aqc-tensor,
  author = {
    James R. Garrison
    and Kate Marshall
    and Ibrahim Shehzad
    and Kevin J. Sung
    and Caleb Johnson
    and Max Rossmannek
    and Bryce Fuller
    and Jennifer R. Glick
    and Albert Akhriev
    and Sergiy Zhuk
    and Niall F. Robertson
  },
  title = {{Qiskit addon: Approximate Quantum Compilation with Tensor Networks}},
  howpublished = {\url{https://github.com/Qiskit/qiskit-addon-aqc-tensor}},
  doi = {10.5281/zenodo.14064353},
  year = {2024}
}

@article{herrmann2022realizing,
  title={Realizing quantum convolutional neural networks on a superconducting quantum processor to recognize quantum phases},
  author={Herrmann, Johannes and Llima, Sergi Masot and Remm, Ants and Zapletal, Petr and McMahon, Nathan A and Scarato, Colin and Swiadek, Fran{\c{c}}ois and Andersen, Christian Kraglund and Hellings, Christoph and Krinner, Sebastian and others},
  journal={Nature communications},
  volume={13},
  number={1},
  pages={4144},
  year={2022},
  publisher={Nature Publishing Group UK London}
}

@article{shen2025robust,
  title={Robust simulations of many-body symmetry-protected topological phase transitions on a quantum processor},
  author={Shen, Ruizhe and Chen, Tianqi and Yang, Bo and Zhong, Yin and Lee, Ching Hua},
  journal={npj Quantum Information},
  volume={11},
  number={1},
  pages={179},
  year={2025},
  publisher={Nature Publishing Group UK London}
}

@article{scheer2025renormalization,
  title={Renormalization-group-based preparation of matrix product states on up to 80 qubits},
  author={Scheer, Moritz and Baiardi, Alberto and Marty, Elisa B{\"a}umer and Wei, Zhi-Yuan and Malz, Daniel},
  journal={arXiv preprint arXiv:2510.24681},
  year={2025}
}

@article{smith2024constant,
  title={Constant-depth preparation of matrix product states with adaptive quantum circuits},
  author={Smith, Kevin C and Khan, Abid and Clark, Bryan K and Girvin, Steven M and Wei, Tzu-Chieh},
  journal={PRX Quantum},
  volume={5},
  number={3},
  pages={030344},
  year={2024},
  publisher={APS}
}

@article{edmunds2025symmetry,
  title={Symmetry-Protected Topological Haldane Phase on a Qudit Quantum Processor},
  author={Edmunds, CL and Rico, E and Arrazola, I and Brennen, GK and Meth, M and Blatt, R and Ringbauer, M},
  journal={PRX Quantum},
  volume={6},
  number={2},
  pages={020349},
  year={2025},
  publisher={APS}
}

@article{smith2022crossing,
  title={Crossing a topological phase transition with a quantum computer},
  author={Smith, Adam and Jobst, Bernhard and Green, Andrew G and Pollmann, Frank},
  journal={Physical Review Research},
  volume={4},
  number={2},
  pages={L022020},
  year={2022},
  publisher={APS}
}

@misc{wang2025direct,
      title={Direct reconstruction of the quantum density matrix elements with classical shadow tomography}, 
      author={Yu Wang},
      year={2025},
      eprint={2505.15243},
      archivePrefix={arXiv},
      primaryClass={quant-ph},
      url={https://arxiv.org/abs/2505.15243}, 
}

@article{nachtergaele2006lieb,
	author = {Nachtergaele, Bruno and Sims, Robert},
	date = {2006/07/01},
	doi = {10.1007/s00220-006-1556-1},
	id = {Nachtergaele2006},
	isbn = {1432-0916},
	journal = {Communications in Mathematical Physics},
	number = {1},
	pages = {119--130},
	title = {Lieb-Robinson Bounds and the Exponential Clustering Theorem},
	url = {https://doi.org/10.1007/s00220-006-1556-1},
	volume = {265},
	year = {2006}}

@article{tenpy2024,
    title={{Tensor network Python (TeNPy) version 1}},
    author={Johannes Hauschild and Jakob Unfried and Sajant Anand and Bartholomew Andrews and Marcus Bintz and Umberto Borla and Stefan Divic and Markus Drescher and Jan Geiger and Martin Hefel and Kévin Hémery and Wilhelm Kadow and Jack Kemp and Nico Kirchner and Vincent S. Liu and Gunnar Möller and Daniel Parker and Michael Rader and Anton Romen and Samuel Scalet and Leon Schoonderwoerd and Maximilian Schulz and Tomohiro Soejima and Philipp Thoma and Yantao Wu and Philip Zechmann and Ludwig Zweng and Roger S. K. Mong and Michael P. Zaletel and Frank Pollmann},
    journal={SciPost Phys. Codebases},
    pages={41},
    year={2024},
    publisher={SciPost},
    doi={10.21468/SciPostPhysCodeb.41},
    url={https://scipost.org/10.21468/SciPostPhysCodeb.41},
}

@article{zhao2024tunable,
  title={Tunable topological phases in nanographene-based spin-1/2 alternating-exchange Heisenberg chains},
  author={Zhao, Chenxiao and Catarina, Gon{\c{c}}alo and Zhang, Jin-Jiang and Henriques, Jo{\~a}o CG and Yang, Lin and Ma, Ji and Feng, Xinliang and Gr{\"o}ning, Oliver and Ruffieux, Pascal and Fern{\'a}ndez-Rossier, Joaqu{\'\i}n and others},
  journal={Nature Nanotechnology},
  volume={19},
  number={12},
  pages={1789--1795},
  year={2024},
  publisher={Nature Publishing Group UK London}
}

@article{mazza2014out,
  title={Out-of-equilibrium dynamics and thermalization of string order},
  author={Mazza, Leonardo and Rossini, Davide and Endres, Manuel and Fazio, Rosario},
  journal={Physical Review B},
  volume={90},
  number={2},
  pages={020301},
  year={2014},
  publisher={APS}
}

@article{mcginley2018topology,
  title={Topology of one dimensional quantum systems out of equilibrium},
  author={McGinley, Max and Cooper, Nigel R},
  journal={arXiv preprint arXiv:1804.05756},
  year={2018}
}

@article{hagymasi2019dynamical,
  title={Dynamical topological quantum phase transitions in nonintegrable models},
  author={Hagym{\'a}si, I and Hubig, Claudius and Legeza, {\"O} and Schollw{\"o}ck, Ulrich},
  journal={Physical Review Letters},
  volume={122},
  number={25},
  pages={250601},
  year={2019},
  publisher={APS}
}

@article{miyake2010quantum,
  title={Quantum computation on the edge of a symmetry-protected topological order},
  author={Miyake, Akimasa},
  journal={Physical review letters},
  volume={105},
  number={4},
  pages={040501},
  year={2010},
  publisher={APS}
}

@article{senthil2015symmetry,
  title={Symmetry-protected topological phases of quantum matter},
  author={Senthil, Todadri},
  journal={Annu. Rev. Condens. Matter Phys.},
  volume={6},
  number={1},
  pages={299--324},
  year={2015},
  publisher={Annual Reviews}
}

@article{daniel2021quantum,
  title={Quantum computational advantage with string order parameters of one-dimensional symmetry-protected topological order},
  author={Daniel, Austin K and Miyake, Akimasa},
  journal={Physical review letters},
  volume={126},
  number={9},
  pages={090505},
  year={2021},
  publisher={APS}
}

@article{song2018observation,
  title={Observation of symmetry-protected topological band with ultracold fermions},
  author={Song, Bo and Zhang, Long and He, Chengdong and Poon, Ting Fung Jeffrey and Hajiyev, Elnur and Zhang, Shanchao and Liu, Xiong-Jun and Jo, Gyu-Boong},
  journal={Science advances},
  volume={4},
  number={2},
  pages={eaao4748},
  year={2018},
  publisher={American Association for the Advancement of Science}
}

@article{de2019observation,
  title={Observation of a symmetry-protected topological phase of interacting bosons with Rydberg atoms},
  author={De L{\'e}s{\'e}leuc, Sylvain and Lienhard, Vincent and Scholl, Pascal and Barredo, Daniel and Weber, Sebastian and Lang, Nicolai and B{\"u}chler, Hans Peter and Lahaye, Thierry and Browaeys, Antoine},
  journal={Science},
  volume={365},
  number={6455},
  pages={775--780},
  year={2019},
  publisher={American Association for the Advancement of Science}
}

@article{cohen2015simulating,
  title={Simulating the Haldane phase in trapped-ion spins using optical fields},
  author={Cohen, I and Richerme, P and Gong, Z-X and Monroe, C and Retzker, A},
  journal={Physical Review A},
  volume={92},
  number={1},
  pages={012334},
  year={2015},
  publisher={APS}
}

@misc{kingma2017adam,
      title={Adam: A Method for Stochastic Optimization}, 
      author={Diederik P. Kingma and Jimmy Ba},
      year={2017},
      eprint={1412.6980},
      archivePrefix={arXiv},
      primaryClass={cs.LG},
      url={https://arxiv.org/abs/1412.6980}, 
}

@article{vatan2004optimal,
  title = {Optimal quantum circuits for general two-qubit gates},
  author = {Vatan, Farrokh and Williams, Colin},
  journal = {Phys. Rev. A},
  volume = {69},
  issue = {3},
  pages = {032315},
  numpages = {5},
  year = {2004},
  month = {Mar},
  publisher = {American Physical Society},
  doi = {10.1103/PhysRevA.69.032315},
  url = {https://link.aps.org/doi/10.1103/PhysRevA.69.032315}
}

@misc{qiskit2024,
      title={Quantum computing with {Q}iskit},
      author={Javadi-Abhari, Ali and Treinish, Matthew and Krsulich, Kevin and Wood, Christopher J. and Lishman, Jake and Gacon, Julien and Martiel, Simon and Nation, Paul D. and Bishop, Lev S. and Cross, Andrew W. and Johnson, Blake R. and Gambetta, Jay M.},
      year={2024},
      doi={10.48550/arXiv.2405.08810},
      eprint={2405.08810},
      archivePrefix={arXiv},
      primaryClass={quant-ph}
}

@article{wallman2016noise,
  title = {Noise tailoring for scalable quantum computation via randomized compiling},
  author = {Wallman, Joel J. and Emerson, Joseph},
  journal = {Phys. Rev. A},
  volume = {94},
  issue = {5},
  pages = {052325},
  numpages = {9},
  year = {2016},
  month = {Nov},
  publisher = {American Physical Society},
  doi = {10.1103/PhysRevA.94.052325},
  url = {https://link.aps.org/doi/10.1103/PhysRevA.94.052325}
}

@article{vandenberg2022model,
  title = {Model-free readout-error mitigation for quantum expectation values},
  author = {van den Berg, Ewout and Minev, Zlatko K. and Temme, Kristan},
  journal = {Phys. Rev. A},
  volume = {105},
  issue = {3},
  pages = {032620},
  numpages = {8},
  year = {2022},
  month = {Mar},
  publisher = {American Physical Society},
  doi = {10.1103/PhysRevA.105.032620},
  url = {https://link.aps.org/doi/10.1103/PhysRevA.105.032620}
}

@article{cai2021multi,
	author = {Cai, Zhenyu},
	date = {2021/05/25},
	doi = {10.1038/s41534-021-00404-3},
	id = {Cai2021},
	isbn = {2056-6387},
	journal = {npj Quantum Information},
	number = {1},
	pages = {80},
	title = {Multi-exponential error extrapolation and combining error mitigation techniques for NISQ applications},
	url = {https://doi.org/10.1038/s41534-021-00404-3},
	volume = {7},
	year = {2021}}

@misc{majumdar2023best,
      title={Best practices for quantum error mitigation with digital zero-noise extrapolation}, 
      author={Ritajit Majumdar and Pedro Rivero and Friederike Metz and Areeq Hasan and Derek S Wang},
      year={2023},
      eprint={2307.05203},
      archivePrefix={arXiv},
      primaryClass={quant-ph},
      url={https://arxiv.org/abs/2307.05203}, 
}

@article{affleck1987rigorous,
  title = {Rigorous results on valence-bond ground states in antiferromagnets},
  author = {Affleck, Ian and Kennedy, Tom and Lieb, Elliott H. and Tasaki, Hal},
  journal = {Phys. Rev. Lett.},
  volume = {59},
  issue = {7},
  pages = {799--802},
  numpages = {0},
  year = {1987},
  month = {Aug},
  publisher = {American Physical Society},
  doi = {10.1103/PhysRevLett.59.799},
  url = {https://link.aps.org/doi/10.1103/PhysRevLett.59.799}
}

@article{vidal2004efficient,
  title={Efficient simulation of one-dimensional quantum many-body systems},
  author={Vidal, Guifr{\'e}},
  journal={Physical review letters},
  volume={93},
  number={4},
  pages={040502},
  year={2004},
  publisher={APS}
}

@book{Nielsen_Chuang_2010, place={Cambridge}, title={Quantum Computation and Quantum Information: 10th Anniversary Edition}, publisher={Cambridge University Press}, author={Nielsen, Michael A. and Chuang, Isaac L.}, year={2010}}

@article{malz2024preparation,
  title = {Preparation of Matrix Product States with Log-Depth Quantum Circuits},
  author = {Malz, Daniel and Styliaris, Georgios and Wei, Zhi-Yuan and Cirac, J. Ignacio},
  journal = {Phys. Rev. Lett.},
  volume = {132},
  issue = {4},
  pages = {040404},
  numpages = {9},
  year = {2024},
  month = {Jan},
  publisher = {American Physical Society},
  doi = {10.1103/PhysRevLett.132.040404},
  url = {https://link.aps.org/doi/10.1103/PhysRevLett.132.040404}
}

@article{rudolph2023decomposition,
doi = {10.1088/2058-9565/ad04e6},
url = {https://doi.org/10.1088/2058-9565/ad04e6},
year = {2023},
month = {nov},
publisher = {IOP Publishing},
volume = {9},
number = {1},
pages = {015012},
author = {Rudolph, Manuel S and Chen, Jing and Miller, Jacob and Acharya, Atithi and Perdomo-Ortiz, Alejandro},
title = {Decomposition of matrix product states into shallow quantum circuits},
journal = {Quantum Science and Technology},
}

@article{pimonpan2022realizing,
	author = {Sompet, Pimonpan and Hirthe, Sarah and Bourgund, Dominik and Chalopin, Thomas and Bibo, Julian and Koepsell, Joannis and Bojovi{\'c}, Petar and Verresen, Ruben and Pollmann, Frank and Salomon, Guillaume and Gross, Christian and Hilker, Timon A. and Bloch, Immanuel},
	date = {2022/06/01},
	doi = {10.1038/s41586-022-04688-z},
	id = {Sompet2022},
	isbn = {1476-4687},
	journal = {Nature},
	number = {7914},
	pages = {484--488},
	title = {Realizing the symmetry-protected Haldane phase in Fermi--Hubbard ladders},
	url = {https://doi.org/10.1038/s41586-022-04688-z},
	volume = {606},
	year = {2022}}

@article{mogerle2025spin,
  title = {Spin-1 Haldane Phase in a Chain of Rydberg Atoms},
  author = {M\"ogerle, J. and Brechtelsbauer, K. and Gea-Caballero, A.T. and Prior, J. and Emperauger, G. and Bornet, G. and Chen, C. and Lahaye, T. and Browaeys, A. and B\"uchler, H.P.},
  journal = {PRX Quantum},
  volume = {6},
  issue = {2},
  pages = {020332},
  numpages = {13},
  year = {2025},
  month = {May},
  publisher = {American Physical Society},
  doi = {10.1103/PRXQuantum.6.020332},
  url = {https://link.aps.org/doi/10.1103/PRXQuantum.6.020332}
}

@article{mishra2021observation,
	author = {Mishra, Shantanu and Catarina, Gon{\c c}alo and Wu, Fupeng and Ortiz, Ricardo and Jacob, David and Eimre, Kristjan and Ma, Ji and Pignedoli, Carlo A. and Feng, Xinliang and Ruffieux, Pascal and Fern{\'a}ndez-Rossier, Joaqu{\'\i}n and Fasel, Roman},
	date = {2021/10/01},
	doi = {10.1038/s41586-021-03842-3},
	id = {Mishra2021},
	isbn = {1476-4687},
	journal = {Nature},
	number = {7880},
	pages = {287--292},
	title = {Observation of fractional edge excitations in nanographene spin chains},
	url = {https://doi.org/10.1038/s41586-021-03842-3},
	volume = {598},
	year = {2021}}

@article{pollmann2010entanglement,
  title = {Entanglement spectrum of a topological phase in one dimension},
  author = {Pollmann, Frank and Turner, Ari M. and Berg, Erez and Oshikawa, Masaki},
  journal = {Phys. Rev. B},
  volume = {81},
  issue = {6},
  pages = {064439},
  numpages = {10},
  year = {2010},
  month = {Feb},
  publisher = {American Physical Society},
  doi = {10.1103/PhysRevB.81.064439},
  url = {https://link.aps.org/doi/10.1103/PhysRevB.81.064439}
}

@article{pollmann2012string,
  title={Symmetry protection of topological phases in one-dimensional quantum spin systems},
  author={Pollmann, Frank and Berg, Erez and Turner, Ari M. and Oshikawa, Masaki},
  journal={Physical Review B},
  volume={85},
  pages={075125},
  year={2012}
}

@article{chen2011classification,
  title={Classification of gapped symmetric phases in one-dimensional spin systems},
  author={Chen, Xie and Gu, Zheng-Cheng and Liu, Zheng-Xin and Wen, Xiao-Gang},
  journal={Physical Review B},
  volume={83},
  pages={035107},
  year={2011}
}

@article{kane2005quantum,
  title={Quantum spin Hall effect in graphene},
  author={Kane, Charles L. and Mele, Eugene J.},
  journal={Physical Review Letters},
  volume={95},
  pages={226801},
  year={2005}
}

@article{caceres2023edge,
  title={Edge-to-edge topological spectral transfer in diamond photonic lattices},
  author={Cáceres-Aravena, Gabriel and Real, Bastián and Guzmán-Silva, Diego and Vildoso, Paloma and Salinas, Ignacio and Amo, Alberto and Ozawa, Tomoki and Vicencio, Rodrigo A.},
  journal={APL Photonics},
  volume={8},
  number={8},
  pages={080801},
  year={2023},
  doi={10.1063/5.0153770}
}

@article{roberts2020symmetry,
  title={Symmetry-protected self-correcting quantum memories},
  author={Roberts, Sam and Bartlett, Stephen D},
  journal={Physical Review X},
  volume={10},
  number={3},
  pages={031041},
  year={2020},
  publisher={APS}
}

@article{else2012symmetry,
  title = {Symmetry-Protected Phases for Measurement-Based Quantum Computation},
  author = {Else, Dominic V. and Schwarz, Ilai and Bartlett, Stephen D. and Doherty, Andrew C.},
  journal = {Phys. Rev. Lett.},
  volume = {108},
  issue = {24},
  pages = {240505},
  numpages = {5},
  year = {2012},
  month = {Jun},
  publisher = {American Physical Society},
  doi = {10.1103/PhysRevLett.108.240505},
  url = {https://link.aps.org/doi/10.1103/PhysRevLett.108.240505}
}

@article{villa2020local,
  title = {Local quench spectroscopy of many-body quantum systems},
  author = {Villa, L. and Despres, J. and Thomson, S. J. and Sanchez-Palencia, L.},
  journal = {Phys. Rev. A},
  volume = {102},
  issue = {3},
  pages = {033337},
  numpages = {12},
  year = {2020},
  month = {Sep},
  publisher = {American Physical Society},
  doi = {10.1103/PhysRevA.102.033337},
  url = {https://link.aps.org/doi/10.1103/PhysRevA.102.033337}
}

@article{hastings2006spectral,
  title={Spectral gap and exponential decay of correlations},
  author={Hastings, Matthew B and Koma, Tohru},
  journal={Communications in mathematical physics},
  volume={265},
  number={3},
  pages={781--804},
  year={2006},
  publisher={Springer}
}

@BIBNOTE{ibmq_systems,
note={IBM Quantum. https://quantum-computing.ibm.com/, 2026}
}

\clearpage

% To update the supplementary information, edit supp.tex, generate a new pdf and overwrite supp.pdf.


\section*{Supplementary material}

# Supplementary Information

This supplementary information is structured as follows. In Sec. S.I we compare the fidelities and CNOT depths achieved by two commonly-used matrix product state (MPS) preparation techniques to those achieved by approximate quantum compiling (AQC). In Sec. S.II we look at the effect of the infidelity introduced by AQC on the entanglement spectra. In Sec. S.III we detail a technique used to verify that the zero-noise extrapolation (ZNE) procedure gives sensible results for circuits with known noiseless expectation values. In Sec. S.IV we show fits for the magnetisation at the edges of the chain, used to extract the correlation length. Finally, in Sec. S.V we give further experimental results for the ground states not shown in the main text.

## S.I. COMPARISON TO OTHER MPS PREPARATION TECHNIQUES

Two commonly used mehods for preparing MPSs as quantum circuits make use of circuits with a staircase, or ladder, structure. The first of these methods is the only exact method of MPS preparation, first introduced in [S1], and explicitly detailed in [S2]. After gauge transforming the MPS into a certain canonical form, each tensor is embedded in a unitary, introducing $|0\rangle$-initialised ancilla qubits as necessary. This results in a staircase of unitaries, where each unitary requires $1 + \lceil \log_2 \chi \rceil$ qubits, where $\chi$ is the local bond dimension. Whilst exact, this method often results in circuits with high circuit depth, due to the complexity of decomposing the large unitaries into single and two-qubit gates. We refer to this method as the "exact ladder (EL)" method.

The second method, related to the EL method, is an approximate method explicitly detailed in [S3], with related methods proposed in [S1, S4]. The method involves iteratively compressing an MPS to bond dimension $\chi = 2$, preparing the compressed MPS exactly as a ladder of two-qubit unitaries using the EL method, applying the inverse of the resulting circuit to the non-compressed MPS, and repeating. The circuit after $L$-iterations of this procedure consist of $L$ ladder layers of two-qubit unitaries which approximately prepares the target state. We refer to this method as the "approximate ladder (AL)" method.

Table S1 compares the fidelity and circuit depths achieved by these methods, using the open source implementation [S5], to those achieved by AQC. For a given ground state, we apply the methods to two compressed versions of the state, one with at least 99.9% fidelity, and the other with at least 98% fidelity. The first compressed target state (top half of Table S1) is the same state as was used as the target for AQC and the second compressed target state (bottom half of Table S1) is intended to allow the other methods a similar infidelity as was obtained by AQC. We note that all fidelities in Table S1 are calculated with respect to the original, uncompressed ground state MPS obtained by DMRG.

| State label | $O_{\frac{1}{2}}$ | | | $E_{\frac{1}{2}}$ | | | $E_{-1}$ | | | $E_{-2}$ | | |
|---|---|---|---|---|---|---|---|---|---|---|---|---|
| Target $\chi$ | 5 | | | 5 | | | 8 | | | 8 | | |
| Target fidelity | 0.999 | | | 0.999 | | | 1.000 | | | 1.000 | | |
| | Fidelity | CNOT depth | CNOT count | Fidelity | CNOT depth | CNOT count | Fidelity | CNOT depth | CNOT count | Fidelity | CNOT depth | CNOT count |
| AQC | 0.989 | 18 | 891 | 0.989 | 21 | 1041 | 0.990 | 21 | 1041 | 0.979 | 39 | 1932 |
| EL | 0.999 | 24,069 | 24,639 | 0.999 | 20,096 | 21,066 | 1.000 | 20,145 | 20,715 | 1.000 | 20,246 | 20,816 |
| AL ($L=1$) | 0.607 | 297 | 297 | 0.597 | 297 | 297 | 0.376 | 297 | 297 | 0.078 | 297 | 297 |
| AL ($L=10$) | 0.803 | 351 | 2970 | 0.798 | 351 | 2970 | 0.644 | 351 | 2970 | 0.343 | 351 | 2970 |
| Target $\chi$ | 4 | | | 4 | | | 4 | | | 7 | | |
| Target fidelity | 0.999 | | | 0.999 | | | 0.994 | | | 0.989 | | |
| | Fidelity | CNOT depth | CNOT count | Fidelity | CNOT depth | CNOT count | Fidelity | CNOT depth | CNOT count | Fidelity | CNOT depth | CNOT count |
| EL | 0.999 | 3368 | 3368 | 0.999 | 3383 | 3383 | 0.994 | 3379 | 3379 | 0.989 | 20,496 | 21,066 |
| AL ($L=1$) | 0.607 | 297 | 297 | 0.597 | 297 | 297 | 0.376 | 297 | 297 | 0.078 | 297 | 297 |
| AL ($L=10$) | 0.806 | 351 | 2970 | 0.801 | 351 | 2970 | 0.648 | 351 | 2970 | 0.317 | 351 | 2970 |

TABLE S1. Comparison of the AQC, EL, and AL compilation techniques. (Top) fidelities, CNOT depths, and CNOT counts of the circuits produced by the three techniques, when using the same target MPS as was used for AQC. Here $L$ refers to the number of ladder layers used for the AL method. (Bottom) results for the EL and AL methods when used with a target MPS compressed to at least 98% fidelity. All fidelities are calculated with respect to the original, uncompressed ground state MPS obtained by DMRG.

Over the four ground states, the AL method only manages to reach fidelities of between 0.078 and 0.806, whilst producing circuits of CNOT depth between 297 and 351. This is compared to fidelities of 0.979 to 0.990 with CNOT depths between 18 and 39 for the AQC method, highlighting the suitability of the brickwork ansatz for these ground states. The EL method manages to reach very high fidelities of 0.979 to 1.000, at the expense of extremely deep circuits, with CNOT depths between 3368 and 24,069: well beyond the capabilities of current generation quantum devices.

We note that other methods for MPS preparation exist, such as ones using a renormalisation-group-based ansatz [S6, S7] or adaptive circuits with mid-circuit measurement and classical feedforward [S8], but we do not expect these approaches to perform favourably for the states under study. The former requires a circuit depth scaling quartically with bond dimension, while the latter is non-deterministic when strict symmetry requirements are not obeyed. We therefore do not conduct an exhaustive comparison with all possible methods for MPS preparation, and narrow our focus to AQC due to its favourable balance of fidelity and circuit depth.

## S.II. THE EFFECT OF AQC APPROXIMATION ON THE ENTANGLEMENT SPECTRUM

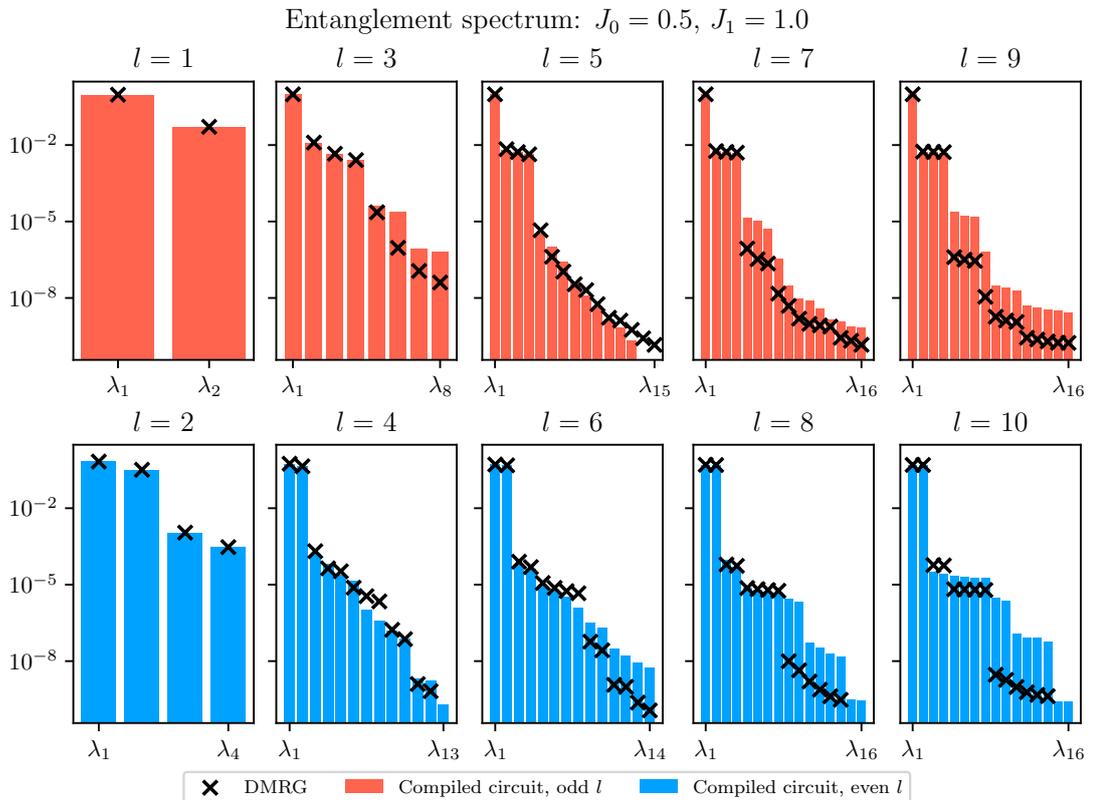

FIG. S1. The effect of AQC approximation error on the entanglement spectrum of the $O_{\frac{1}{2}}$ ground state. The red (blue) bars show up to the sixteen largest eigenvalues, $\lambda_i$, of the $l$-site reduced density matrix obtained by cutting $J_0$ ($J_1$) bonds, calculated via MPS simulation of the AQC compiled circuit. The black crosses show the corresponding values obtained from DMRG. We discard any values below $10^{-10}$.

In the experiments section of the main text, we present experimental results for the entanglement spectrum of the $O_{\frac{1}{2}}$ and $E_{-2}$ ground states obtained using `ibm_pittsburgh` (Fig. 5 in the main text). We compare the results to the corresponding spectrum obtained by DMRG. However, we do not show the entanglement spectrum obtained by MPS simulation of the AQC compiled circuit, since is is almost indistinguishable from the DMRG spectrum on the scale of the plot. In this section, we look at the effect of the AQC approximation error on the entanglement spectrum.

Fig. S1 and Fig. S2 show the entanglement spectrum for cuts of length $l = 1$ to 10 for the $O_{\frac{1}{2}}$ and $E_{-2}$ ground states respectively. The red and blue bars show the values obtained via classical MPS simulation of the AQC compiled circuit, and the black crosses show those obtained from DMRG. We observe that the entanglement spectrum values










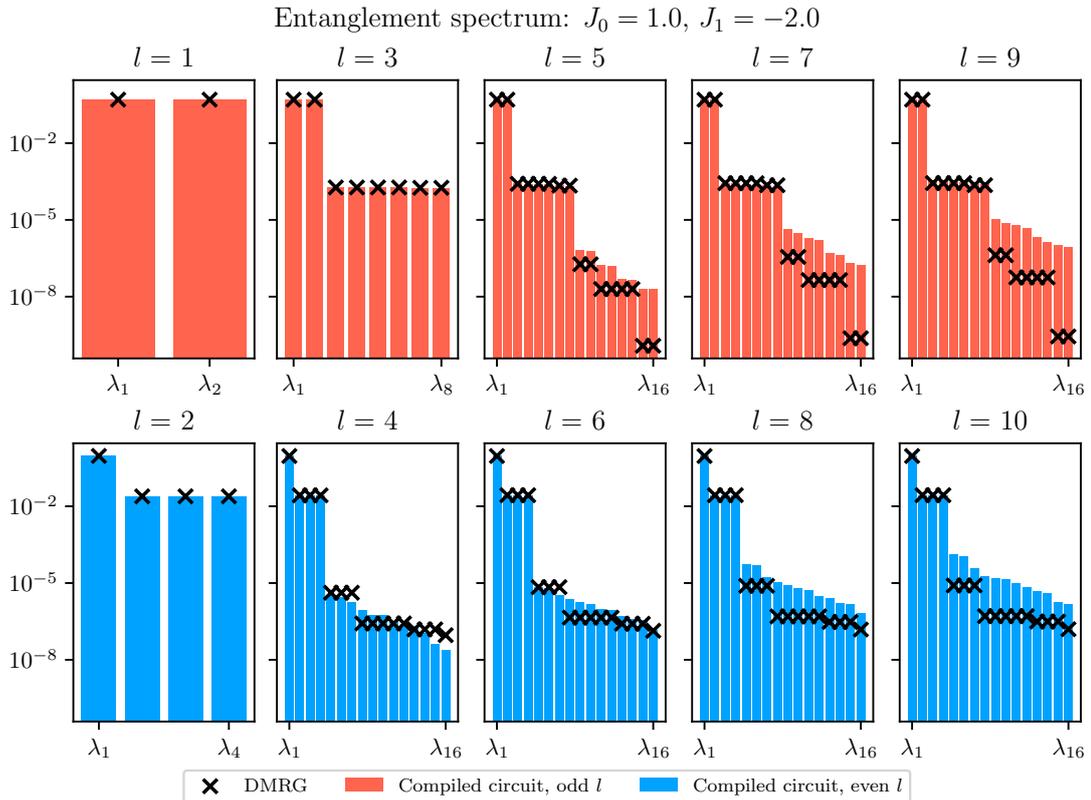

FIG. S2. The effect of AQC approximation error on the entanglement spectrum of the $E_{-2}$ ground state. The red (blue) bars show up to the sixteen largest eigenvalues, $\lambda_i$, of the $l$-site reduced density matrix obtained by cutting $J_0$ ($J_1$) bonds, calculated via MPS simulation of the AQC compiled circuit. The black crosses show the corresponding values obtained from DMRG. We discard any values below $10^{-10}$.

from DMRG and from the AQC compiled circuit are in good agreement down to a scale of around $10^{-4}$. Below $10^{-4}$, the values are not captured well by AQC, and the signature symmetry-protected topological (SPT) degeneracies break down.

## S.III. IDENTITY CIRCUIT ZNE

An important consideration when using the ZNE error mitigation technique [S9] is the choice of noise factors. If the chosen noise factors cover a range that is too small, there would be very little variation in the measured expectation value, and fitting procedures would be unstable. On the other hand, if the range of noise factors is too large, the measured expectation values would quickly lose any information from the circuit. Following the idea of using circuits with known noiseless expectation values for calibrating the noise factors mentioned in [S10], we use a technique using so called "identity circuits" to verify that the chosen range of noise factors used is suitable.

Our AQC circuits consist of a brickwork structure of layers of SU(4) universal two-qubit unitaries, decomposed using the Cartan decomposition [S11]. The parameters of each SU(4) unitary are determined by the AQC procedure. We can construct a circuit with the same structure as the AQC circuit, but which corresponds to the identity operation by choosing parameters such that each SU(4) unitary corresponds to the identity. When transpiling the circuit to the device with the Qiskit [S12] transpiler, using optimisation level 1 ensures that the transpiler does not remove the identity SU(4) unitaries. With such a circuit, in the absence of noise, the expectation value of single-qubit Pauli-Z operators is: $\langle \hat{Z}_i \rangle = 1$. Ideally, performing the ZNE procedure with such a circuit, we expect a) the unmitigated (noise factor 1) value should be close to 1 b) this value should decay fairly monotonically and smoothly over the range of noise factors, and c) when extrapolated to zero noise, the value should be close to 1. If we observe that the unmitigated value is close to zero, or the values behave erratically, then the circuit is likely too deep for the device, or the range of noise factors is too large. Another possibility is that the chosen circuit layout on the quantum device

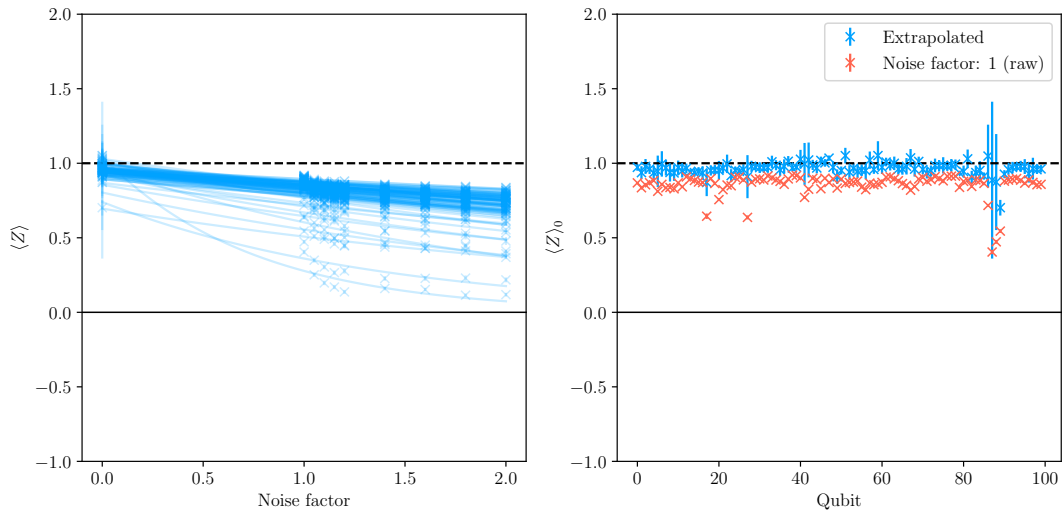

FIG. S3. Results of the identity ZNE procedure for the experiment shown in Fig. 3, using `ibm_pittsburgh`. (Left) Pauli-Z expectation value, $\langle \hat{Z}_i \rangle$ as a function of noise factor, for all qubits ($i = 0, ..., 99$). (Right) Extrapolated value (blue, noise factor 0) and unmitigated value (red, noise factor 1) as a function of virtual qubit index.

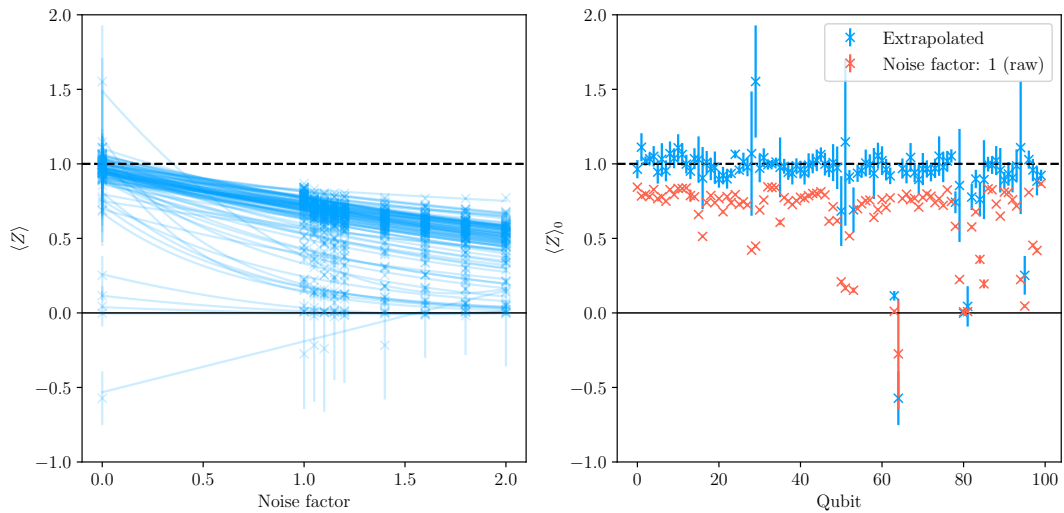

FIG. S4. Example of the effect of bad qubits on the identity ZNE procedure, using `ibm_fez`. (Left) Pauli-Z expectation value, $\langle \hat{Z}_i \rangle$ as a function of noise factor, for all qubits ($i = 0, ..., 99$). (Right) Extrapolated value (blue, noise factor 0) and unmitigated value (red, noise factor 1) as a function of virtual qubit index.

could include one or more particularly noisy qubits. In this case, such a qubit should stand out when performing ZNE with the identity circuit.

Immediately before executing the AQC circuits in our experiments, we execute an identity circuit as described above, and perform the ZNE procedure. We use the same circuit layout on the quantum device as is used for the real experiment. Fig. S3 shows the results of such an experiment, performed on `ibm_pittsburgh` immediately prior to the experiment shown in Fig. 3 in the main text, using an identity circuit with the same structure and layout as the $E_{-2}$ compiled circuit. We observe that the unmitigated $\langle \hat{Z} \rangle$ expectation values mostly lie between 0.8 and 0.9, with the lowest value around 0.4. The behaviour of the expectation values with increasing noise factor is smooth, with no erratic jumps, and the extrapolated values all lie close to the true value of 1. There are also no qubits which behave significantly differently from the rest. We conclude that the ZNE procedure with the identity circuit is working as intended, and we can be confident that: the circuit is not too deep for the device, the range of noise factors is suitable,

and the transpiler did not choose a layout with any particularly noisy qubits. From this, we infer that the same holds for the real experiment with the compiled circuit.

Fig. S4 shows the effect of bad qubits on the identity ZNE procedure, using `ibm_fez`. Here, multiple qubits exhibit expectation values which are either all close to zero, or oscillate erratically as a function of noise factor, and result in poor extrapolations. Qubit 64 appears to be particularly bad, with a negative extrapolated expectation value. In this case, we conclude that the chosen circuit layout included particularly error-prone qubits, and discard the experiment.

## S.IV. EDGE MAGNETISATION DECAY

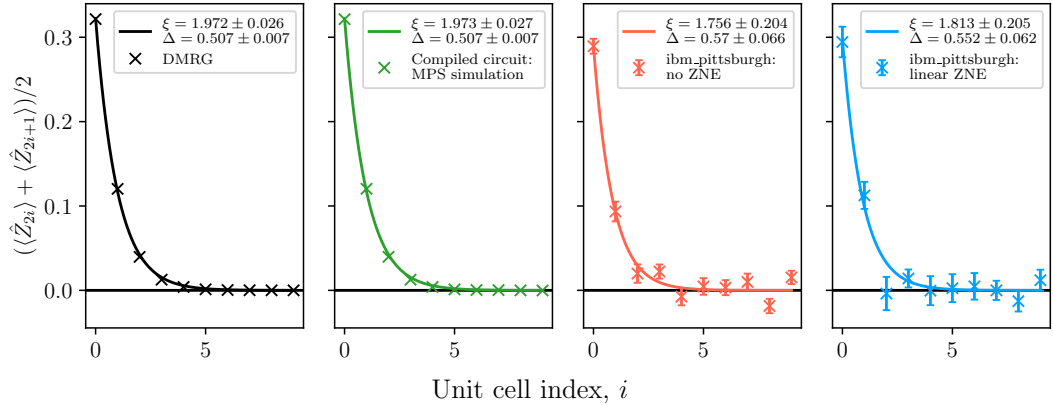

FIG. S5. Fits of the magnetisation of each two-site unit cell, $S_1^z = (\langle \hat{Z}_{2i} \rangle + \langle \hat{Z}_{2i+1} \rangle)/2$, from the left end of the chain, to the functional form: $S_1^z \propto e^{-x/\xi_1}$. The blue (red) crosses show the expectation values obtained using `ibm_pittsburgh` with (without) linear ZNE. The black and green crosses show the corresponding values obtained classically from DMRG and from MPS simulation of the compiled circuit, respectively.

In Fig. 4b in the main text we show the single-site magnetisation for the twenty sites closest to each end of the chain, for the $O_{\frac{1}{2}}$ ground state, obtained using `ibm_pittsburgh`. As detailed in the results discussion, we fit the magnetisation of each two-site unit cell, $S_1^z = (\langle \hat{Z}_{2i} \rangle + \langle \hat{Z}_{2i+1} \rangle)/2$, from the left end of the chain, to the functional form: $S_1^z \propto e^{-x/\xi_1}$, and use this to extract the correlation length, $\xi = 2\xi_1$. Fig.S5 shows the results of the fitting procedure. We obtain $\xi = 1.97 \pm 0.03$ (DMRG and AQC), $1.76 \pm 0.20$ (hardware, no ZNE), and $1.81 \pm 0.21$ (hardware, linear ZNE).

## S.V. FURTHER EXPERIMENTAL RESULTS

Fig. 3 and Fig. 4 in the experiments section of the main text shows the string order parameters as a function of string length, for the $E_{-2}$ and $O_{\frac{1}{2}}$ ground states respectively, as measured using `ibm_pittsburgh`. Here, Figures S6 and S7 show the string order parameters, $S_{l,s}^{O}$ (left) and $S_{l,s}^{E}$ (right), for the $E_{\frac{1}{2}}$, and $E_{-1}$ ground states, respectively. The blue and red crosses show the expectation values as measured using `ibm_pittsburgh` with and without using ZNE, respectively. The black and green crosses show the corresponding quantities obtained classically using DMRG and MPS simulation of the compiled circuits, respectively. The values in the plot are averaged over five sections of the chain, corresponding to $s = 20, 30, 40, 50,$ and $60$. The results are consistent with those of the $E_{-2}$ ground state: $S_{l,s}^{O}$ decays to zero, consistent with the absence of odd string order, and $S_{l,s}^{E}$ remains non-zero even up to $l = 20$, indicating the presence of even string order.

Fig. 5 shows the entanglement spectra of the $O_{\frac{1}{2}}$ and $E_{-2}$ ground states, obtained using `ibm_pittsburgh`, for cuts of length $l = 1$ to $6$. Here, Figures S8 and S9 show the entanglement spectra of the $E_{\frac{1}{2}}$ and $E_{-1}$ ground states, respectively, for cuts of length $l = 1$ to $4$. The red (blue) bars show the eight largest eigenvalues, $\lambda_1 \geq ... \geq \lambda_8$, of the reduced density matrix obtained by cutting $J_0$ ($J_1$) bonds. The black crosses show the corresponding values obtained from DMRG. The results are consistent with those of the $E_{-2}$, with spectra resulting from cuts to $J_0$ bonds having two dominant eigenvalues, and those from cuts to $J_1$ bonds having only a single dominant eigenvalue.

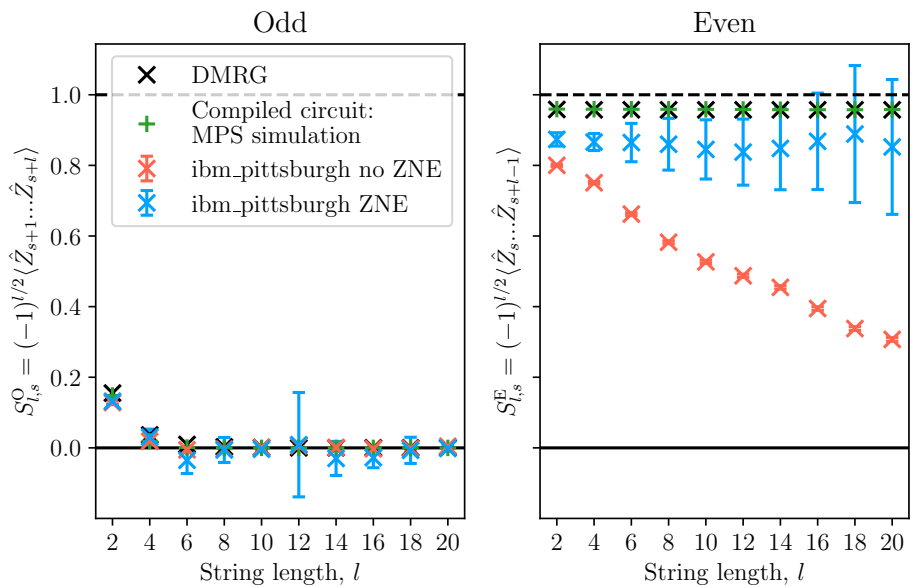

FIG. S6. String order parameters for the $E_{\frac{1}{2}}$ ground state, averaged over five sections of the chain, obtained using `ibm_pittsburgh`. The blue and red crosses show the expectation values as measured using `ibm_pittsburgh` with and without using ZNE, respectively. The error bars represent the standard error on the mean (red) and the uncertainty of the fit evaluated at a noise factor of zero (blue), after propagation for the average value using the relation for a sum. The black and green crosses show the corresponding quantities obtained classically using DMRG and MPS simulation of the compiled circuits, respectively. The black dashed line represents the maximum possible value of 1.

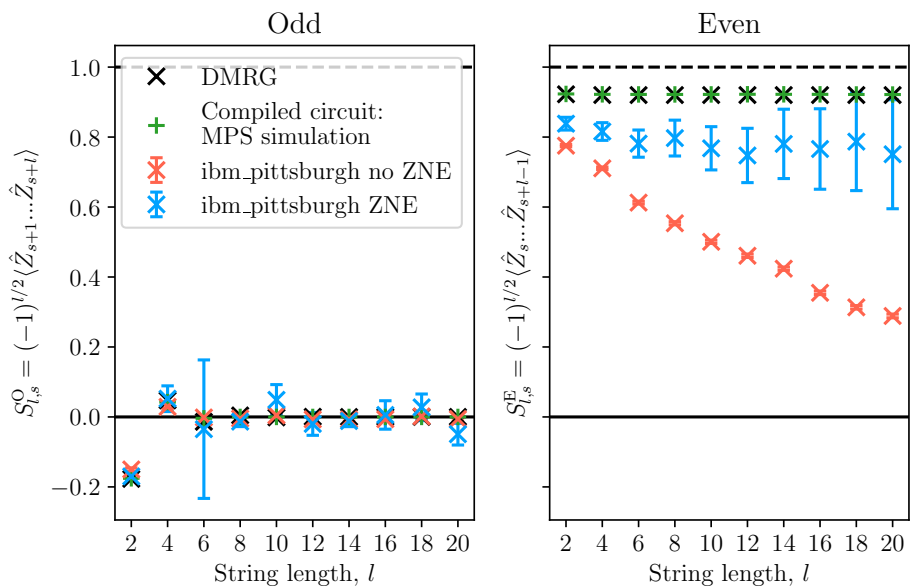

FIG. S7. String order parameters for the $E_{-1}$ ground state, averaged over five sections of the chain, obtained using `ibm_pittsburgh`. The meanings of all elements of the plot are the same as in Fig. S6.

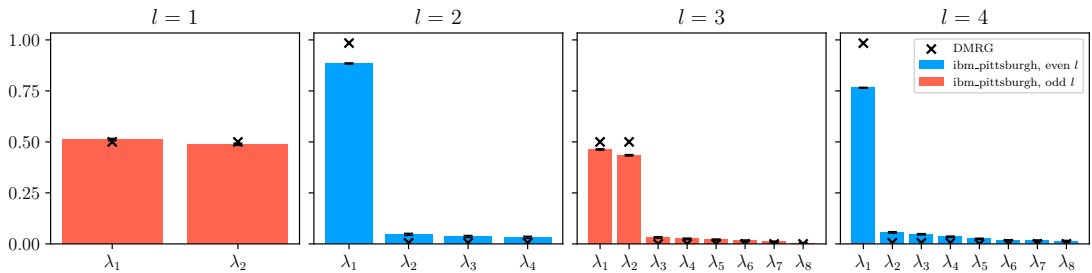

FIG. S8. Entanglement spectra of the $E_{\frac{1}{2}}$ ground state for cuts of up to $l = 4$ sites, executed on `ibm_pittsburgh`. The red (blue) bars show the eight largest eigenvalues, $\lambda_1 \geq ... \geq \lambda_8$, of the reduced density matrix obtained by cutting $J_0$ ($J_1$) bonds. The black crosses show the corresponding values obtained from DMRG. The error bars are obtained via a bootstrapping procedure, see Methods Sec. C 2 in the main text for more details.

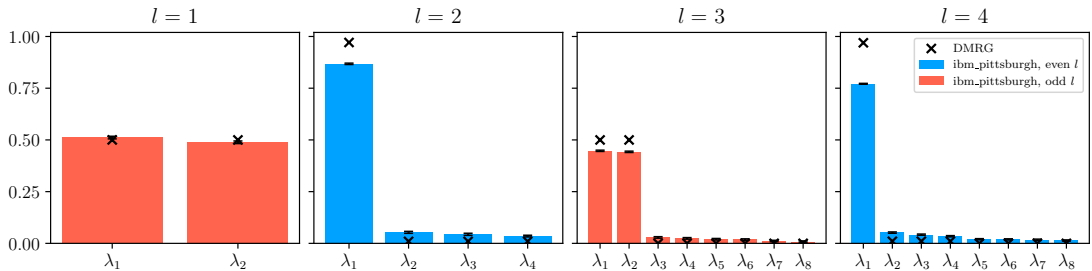

FIG. S9. Entanglement spectra of the $E_{-1}$ ground state for cuts of up to $l = 4$ sites, executed on `ibm_pittsburgh`. The meanings of all elements of the plot are the same as in Fig. S8.

[S1] C. Schön, E. Solano, F. Verstraete, J. I. Cirac, and M. M. Wolf, Sequential generation of entangled multiqubit states, Physical review letters **95**, 110503 (2005).
[S2] S.-H. Lin, R. Dilip, A. G. Green, A. Smith, and F. Pollmann, Real- and imaginary-time evolution with compressed quantum circuits, PRX Quantum **2**, 010342 (2021).
[S3] S.-J. Ran, Encoding of matrix product states into quantum circuits of one-and two-qubit gates, Physical Review A **101**, 032310 (2020).
[S4] C. Schön, K. Hammerer, M. M. Wolf, J. I. Cirac, and E. Solano, Sequential generation of matrix-product states in cavity qed, Phys. Rev. A **75**, 032311 (2007).
[S5] D. A. Millar, G. W. Pennington, N. T. M. Siow, and S. J. Thomson, qiskit-community/mps-to-circuit (2025).
[S6] M. Scheer, A. Baiardi, E. B. Marty, Z.-Y. Wei, and D. Malz, Renormalization-group-based preparation of matrix product states on up to 80 qubits, arXiv preprint arXiv:2510.24681 (2025).
[S7] D. Malz, G. Styliaris, Z.-Y. Wei, and J. I. Cirac, Preparation of matrix product states with log-depth quantum circuits, Phys. Rev. Lett. **132**, 040404 (2024).
[S8] K. C. Smith, A. Khan, B. K. Clark, S. Girvin, and T.-C. Wei, Constant-depth preparation of matrix product states with adaptive quantum circuits, PRX Quantum **5**, 030344 (2024).
[S9] Z. Cai, Multi-exponential error extrapolation and combining error mitigation techniques for nisq applications, npj Quantum Information **7**, 80 (2021).
[S10] R. Majumdar, P. Rivero, F. Metz, A. Hasan, and D. S. Wang, Best practices for quantum error mitigation with digital zero-noise extrapolation (2023), arXiv:2307.05203 [quant-ph].
[S11] F. Vatan and C. Williams, Optimal quantum circuits for general two-qubit gates, Phys. Rev. A **69**, 032315 (2004).
[S12] A. Javadi-Abhari, M. Treinish, K. Krsulich, C. J. Wood, J. Lishman, J. Gacon, S. Martiel, P. D. Nation, L. S. Bishop, A. W. Cross, B. R. Johnson, and J. M. Gambetta, Quantum computing with Qiskit (2024), arXiv:2405.08810 [quant-ph].



%TC:endignore
\end{document}